\documentclass[aps,reprint,prx,letterpaper,floatfix,nobalancelastpage,notitlepage,superscriptaddress]{revtex4-1}
\usepackage{amsmath}
\usepackage{amsfonts}
\usepackage{graphicx}
\usepackage{bm}
\usepackage{color}
\usepackage{braket}
\usepackage[utf8]{inputenc}

\usepackage{color}

\begin{document}

\abovedisplayskip=6pt
\abovedisplayshortskip=6pt
\belowdisplayskip=6pt
\belowdisplayshortskip=6pt

\title{Unidirectional Quantum Transport in Optically Driven \textit{V}-type Quantum Dot Chains}
\author{Oliver Kaestle}
\email{o.kaestle@tu-berlin.de}
\affiliation{Technische Universit\"at Berlin, Institut f\"ur Theoretische Physik, Nichtlineare Optik und Quantenelektronik, Hardenbergstraße 36, 10623 Berlin, Germany}
\author{Emil Vosmar Denning}
\affiliation{DTU Fotonik, Department of Photonics Engineering, Technical University of Denmark, \O rsteds Plads, DK-2800 Kongens Lyngby, Denmark}
\author{Jesper M\o rk}
\affiliation{DTU Fotonik, Department of Photonics Engineering, Technical University of Denmark, \O rsteds Plads, DK-2800 Kongens Lyngby, Denmark}
\author{Andreas Knorr}
\affiliation{Technische Universit\"at Berlin, Institut f\"ur Theoretische Physik, Nichtlineare Optik und Quantenelektronik, Hardenbergstraße 36, 10623 Berlin, Germany}
\author{Alexander Carmele}
\affiliation{Technische Universit\"at Berlin, Institut f\"ur Theoretische Physik, Nichtlineare Optik und Quantenelektronik, Hardenbergstraße 36, 10623 Berlin, Germany}
\date{\today}

\begin{abstract}
We predict a mechanism for achieving complete population inversion in a continuously driven InAs/GaAs semiconductor quantum dot featuring \textit{V}-type transitions. This highly nonequilibrium steady state is enabled by the interplay between \textit{V}-type interband transitions and a non-Markovian decoherence mechanism, introduced by acoustic phonons.
The population trapping mechanism is generalized to a chain of coupled emitters. Exploiting the population inversion, we predict unidirectional excitation transport from one end of the chain to the other without external bias, independent of the unitary interdot coupling mechanism.
\end{abstract}

\maketitle

\newpage

\section{Introduction}

Utilizing and controlling non-Markovian effects is of central importance for the development of quantum optical devices and quantum information technology~\cite{Liu2011, Liu2013, Reich2015, Wolf2008, Chan2014, Jing2010, Kaer2013}. Specifically, the interplay of dissipative processes and phonon-assisted coherence in semiconductor nanostructures such as quantum dots has been established as a growing field of research~\cite{Kozlov2006, Stace2005, Kaer2010, Weiler2012, Denning2020, Stock2011, Hughes2011, Thoma2016, Su2013, Kabuss2011, kabuss2011inductive, Naesby2008, Reigue2017, kreinberg2018quantum}.
For instance, non-Markovian system-reservoir interactions have been shown to create population inversion up to a certain degree in such emitters~\cite{Ardelt2014, Glaessl2011, Glaessl2013, Barth2016fast, Barth2016path}, making them promising candidates for single-exciton light sources on an integrated chip basis. Approaches to achieve inversion range from adiabatic rapid passage in quantum emitters~\cite{Wu2011} to setups of quantum dots coupled to metal nanoparticles~\cite{Paspalakis2013} and cavities~\cite{Roy2011,Hughes2013,Metelmann2015}. Most recently, population inversion was achieved in a single InAs/GaAs quantum dot via pulsed excitations tuned within the exciton phonon sideband, where a phonon-mediated thermalization of the optically dressed states enabled the inversion~\cite{Quilter2015,Thomas2020}. Already, these examples show the importance of electron-phonon interaction as a useful resource to achieve tailored optical excitations.

In this work, we describe a new transport effect in quantum dots enabled by non-Markovian electron-phonon interaction: We predict unidirectional transport in quantum dots with \textit{V}-type transitions continuously driven by a single laser field~\cite{Houmark2009, Barettin2009, Wang2005}. As a result of resonant excitation of an electronic state in a quantum dot and its interaction with a lower-energy state, non-reciprocal phonon-assisted energy transfer takes place in single quantum dots and quantum dot chains.
Specifically, the non-Markovian description of the environment allows for information backflow, such that the structured reservoir can support excitation transfer within the system~\cite{Breuer2002,Vega2017, Carmele2019}.

\begin{figure}[b]
\centering
\includegraphics[width=0.9\linewidth]{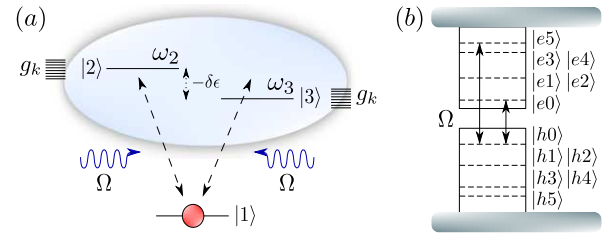}
\caption{(a) Sketch of a \textit{V}-type emitter model. The two excited states have an energy detuning $\delta \epsilon = \hbar \left( \omega_3 - \omega_2 \right)$. A continuous laser field drives the transitions between ground and excited states at a Rabi frequency $\Omega$. The upper levels are coupled to a structured 3D acoustic phonon reservoir via coupling elements $g_{\bm{k}}$. The initial state contains one electron in state $\ket{1}$. (b) Schematic quantum dot energy level structure of InAs sandwiched between two slabs of GaAs, featuring \textit{V}-type transitions between electron and hole states.}
\label{fig:vsystem}
\end{figure}

(i) For a single quantum dot, this results in the preparation of a highly nonequilibrium steady state. For certain parameters, e.g. depending on coupling and driving strength, complete population inversion can be achieved in the detuned excited state. In difference to previously reported mechanisms~\cite{Ardelt2014, Glaessl2011, Glaessl2013,  Barth2016fast, Wu2011, Hughes2013}, the inversion does not rely on cavity interactions or pulsed excitations and is found to be robust for a wide range of coupling and driving strengths.

(ii) To provide an application of the observed population inversion for excitation transport, several quantum dots are combined to a chain-like spatial distribution. Exploiting the population inversion in the red-detuned excited states, we predict unidirectional excitation transport from one end of the quantum dot chain to the other.
Unidirectional energy transfer in the form of electronic excitation transport has been widely investigated in semiconductor nanostructures~\cite{Beenakker1991, Crooker2002, Machida2003, Sangu2003, Lu2007, Chandler2007, Iotti2005, Richter2006}, molecules~\cite{Metivier2007, Leitner2015} and biological systems~\cite{Cupellini2018}. Based on well established transfer protocols, e.g. enabled by an external bias~\cite{Shangguan2001}, it was shown that non-Markovian system-reservoir interactions can influence non-reciprocal quantum transport~\cite{Zheng2009} and increase its efficiency~\cite{Rebentrost2009}.
In contrast to these earlier investigations, here we provide an example in which unidirectional quantum transport results without the application of an external potential, but solely from the interplay between coherence and incoherence in the system itself. Due to the robustness of the underlying mechanism, the transport withstands the influence of perturbations such as radiative decay phenomena and intraband phonon couplings, and emerges regardless of the specifically featured interdot coupling mechanism.

For the theoretical description of the system, two independent approaches are employed. Firstly, we calculate the system dynamics using a polaron master equation in second order perturbation theory with respect to the polaron Hamiltonian~\cite{Breuer2002, Wurger1998, Wilson2002, Weiler2012, Manson2016, Lee2012, McCutcheon2010, Jang2008, Kolli2011, Chang2013, Carmele2019}. Secondly, to gain insight into the key interaction processes enabling the population inversion mechanism and the specific channels of information backflow between system and reservoir, we reproduce the dynamics in the Heisenberg picture and then use it to unravel the dynamics.


\section{Hamiltonian and equations of motion}

We investigate the \textit{V}-type emitter model system shown in Fig.~\ref{fig:vsystem}(a), consisting of a single ground state $\ket{1}$ and two excited states $\ket{2}$ and $\ket{3}$, with an energy detuning $\delta \epsilon = \hbar(\omega_3 - \omega_2)$ between them. The considered \textit{V}-type transitions are driven by a single continuous laser field with Rabi frequency $\Omega$.
A possible realization of this system is constituted by conical InAs quantum dots surrounded by GaAs bulk material~\cite{Dachner2010}. The corresponding energy level structure is shown schematically in Fig.~\ref{fig:vsystem}(b). The electronic structure, within the envelope wave function and effective mass approximation, features six confined electron and hole states denoted by $\ket{e0},\ldots,\ket{e5}$ and $\ket{h0},\ldots,\ket{h5}$, respectively~\cite{Melnik2003, Ahn2005, Houmark2009, Barettin2009}. The rotational symmetry of the conical structure results in degeneracy of the first and second as well as the third and fourth states in both bands~\cite{Houmark2009, Barettin2009}. Therefore, for the realization of a \textit{V}-type transition by an external field, the dipole selection rules allow for two interband transitions from the heavy hole state $\ket{h0}$ to the electronic states $\ket{e0}$ and $\ket{e5}$~\cite{Houmark2009} [cf. Fig.~\ref{fig:vsystem}(b)], corresponding to the two optical transitions in our considered three level model system.
When choosing the laser frequency as a rotating frame and applying the rotating wave approximation, the Hamiltonians for the quantum dot $H_{el}$ and the electron-light coupling $H_l$ are given by
\begin{align}
H_{el} &= \hbar \left( \Delta_2 \sigma_{22} + \Delta_3 \sigma_{33} \right), \\
H_{l} &= \hbar \Omega (\sigma_{12} + \sigma_{13} + \mathrm{H.c.}),
\label{eq:H_el_l}
\end{align}
with $\sigma_{ij}= \ket{i} \bra{j}$. $\Delta_i=\omega_i-\omega_L$ denotes the detuning between the $i$-th excited state frequency and the incident laser frequency, yielding an energy detuning $\delta \epsilon = \hbar (\Delta_3 - \Delta_2)$ between the excited states, and $\Omega$ is the real valued, slowly varying envelope of the Rabi frequency of transitions between ground and excited states.

To account for decoherence effects caused by the embedding GaAs material, a structured phonon reservoir is coupled to the electron states $\ket{2}$ and $\ket{3}$ of the quantum emitter [cf. Fig.~\ref{fig:vsystem}(b)]. In this work, we focus on a generic model of diagonal electron-phonon interactions $H_{el,ph}$ for 3D bulk phonons of the form~\cite{Abrikosov1965,Mahan2000}
\begin{equation}
H_{el,ph} = \hbar \! \int \! \mathrm{d}^3 k \ g_{\bm{k}} \left( \sigma_{22} + \sigma_{33} \right) \left( r_{\bm{k}}^\dagger + r_{\bm{k}} \right),
\label{eq:H_elph}
\end{equation}
with bosonic annihilation (creation) operators $r_{\bm{k}}^{(\dagger)}$. The wave vector ($\bm{k}$) dependence of the acoustic phonon coupling element for GaAs is given by $g_{\bm{k}}^{ii} = \sqrt{ \hbar k/(2 \rho c_s )} D_i \exp [ - \hbar k^2/(4 m_i \omega_i) ]$, with the transition coupling elements for the two excited states prescribed by $g_{\bm{k}}^2 := g_{\bm{k}}^{22} - g_{\bm{k}}^{11}$ and $g_{\bm{k}}^3 := g_{\bm{k}}^{33} - g_{\bm{k}}^{11}$~\cite{Carmele2019, Foerstner2003, CarmeleMilde2013}.
Here, $c_s$ denotes the sound velocity, $D_i$ are the deformation potentials, $m_i$ refer to the effective masses, $\hbar \omega_i$ are the confinement energies and $\rho$ denotes the mass density of GaAs, respectively. For simplicity we assume $g_{\bm{k}}^2=g_{\bm{k}}^3=g_{\bm{k}}$.
However, we stress that the following results describe a generic effect that is robust with respect to the specific form of the coupling element $g_k$, as long as phonon-assisted resonances allow for Stokes processes. Alternative forms of $g_k$ such as Gaussian and Lorentzian-shaped dependencies have been tested and lead to qualitative and quantitative comparable results.
For the moment, we neglect phonon-induced intraband couplings $\sim g_{\bm{k}}^{23} \sigma_{23}$ and radiative decay, as our key results can be shown to be robust against their presence.
Together with the free phonon evolution $H_{ph} = \hbar \! \int \! \mathrm{d}^3 k \  \omega_k r_{\bm{k}}^\dagger r_{\bm{k}}$, where $\omega_k=c_s |{\bm{k}}|$ denotes the acoustic phonon frequencies, we arrive at the full open system Hamiltonian $H = H_{el} + H_{l} + H_{ph} + H_{el,ph}$.
For all following calculations, we choose the parameters $T=4\,$K, $\delta \epsilon=-1.0\,$meV and $\Omega=0.1\,\mathrm{ps}^{-1}$ as temperature, detuning energy and Rabi frequency of the driving field, respectively.

For the numerical evaluation, we employ the standard second-order perturbative polaron master equation~\cite{Breuer2002, Wurger1998, Wilson2002, Weiler2012, Manson2016, Lee2012, McCutcheon2010, Jang2008, Kolli2011, Chang2013, Carmele2019},
\begin{align}
&\partial_t \rho_S(t) = - \dfrac{i}{\hbar} \left[H_{p,0}, \rho_S (t) \right] \nonumber \\
&- \dfrac{1}{\hbar^2} \int_0^t d\tau \mathrm{tr}_B \{ \left[ H_{p,I}, \left[ {H}_{p,I} (-\tau), \rho_S (t) \otimes \rho_B \right] \right] \},
\label{eq:redfield}
\end{align}
where $H_{p,0}$ and $H_{p,I}$ denote the polaron-transformed homogeneous and system-reservoir interaction Hamiltonians, respectively. They are given by
\begin{align}
H_{p,0}/ \hbar = \sum_{i=2,3} \left[ \bar{\Delta}_{i} \sigma_{ii} + \bar{\Omega} (\sigma_{1i} + \sigma_{i1}) \right]
\end{align}
and
\begin{align}
H_{p,I} / \hbar = \sum_{i=2,3} \big[ &(\sigma_{1i} + \sigma_{i1}) \left( \Omega \cosh{(R^\dagger - R)} - \bar{\Omega} \right) \nonumber \\
&+ \Omega (\sigma_{i1} - \sigma_{1i}) \sinh{(R^\dagger - R)} \big],
\end{align}
with collective bosonic operators $R^{(\dagger)} = \int \mathrm{d}^3 k \, (g_{\bm{k}}/\omega_k) r_{\bm{k}}^{(\dagger)}$, a polaron-shifted detuning $\bar{\Delta}_i$ and $\bar{\Omega}= \Omega \exp \left[ -1/2 \int \mathrm{d} \bm{k} \  g_{\bm{k}}/\omega_k \coth \left( \hbar \omega_k / (2k_BT) \right) \right]$.
Here, a Franck-Condon renormalization was applied by introducing $\bar{\Omega}$, such that $\mathrm{tr}_B \{ \left[ H_{p,I}, \rho(t) \right] \}=0$. The time-reversed unitary evolution of the interaction $H_{p,I}(-\tau)=U^\dagger(-\tau,0)H_{p,I}U(-\tau,0)$ is governed by $H_{p,0}$ via $U(t,0)=\exp(-i/\hbar H_{p,0} t)$.
The resulting equations of motion for the density matrix elements $\rho_{mn}$ read
\begin{align}
&\partial_t \rho_{mn} = \sum_{i=2,3} \Big[ i \bar{\Delta}_i \left( \rho_{mi} \delta_{ni} - \rho_{in} \delta_{mi} \right) \nonumber \\
&+ i \bar{\Omega} (\rho_{m1} \delta_{ni} + \rho_{mi} \delta_{n1} - \rho_{in} \delta_{m1} - \rho_{1n} \delta_{mi} ) \Big] \nonumber \\
& - \dfrac{\bar{\Omega}^2}{\hbar^2} \sum_{i,j=2,3} \int_0^t d\tau \bra{m} \chi^{ij}(\tau) \ket{n},
\label{eq:pmeq_eqmotion}
\end{align}
with
\begin{align}
&\bra{m} \chi^{ij}(\tau) \ket{n} \nonumber \\
&= \sum_{q=1}^3 \Big( \rho_{qn} \Big[ \left( G_+(\tau) X_{+,j}^{iq}(-\tau) -i G_-(\tau) X_{-,j}^{iq}(-\tau) \right) \delta_{m1} \nonumber \\
& + \left( G_+(\tau) X_{+,j}^{1q}(-\tau) + iG_-(\tau) X_{-,j}^{1q}(-\tau) \right) \delta_{mi} \Big] \nonumber \\
&+ \rho_{mq} \Big[ \left( G_+^*(\tau) X_{+,j}^{q1}(-\tau) -i G_-^*(\tau) X_{-,j}^{q1}(-\tau) \right) \delta_{ni} \nonumber \\
& + \left( G_+^*(\tau) X_{+,j}^{qi}(-\tau) + iG_-^*(\tau) X_{-,j}^{qi}(-\tau) \right) \delta_{n1} \Big] \nonumber \\
&+\rho_{q1} \Big[ - G_+(\tau) X_{+,j}^{mq}(-\tau) + i G_-(\tau) X_{-,j}^{mq}(-\tau) \Big] \delta_{ni} \nonumber \\
&+\rho_{qi} \Big[ - G_+(\tau) X_{+,j}^{mq}(-\tau) - i G_-(\tau) X_{-,j}^{mq}(-\tau) \Big] \delta_{n1} \nonumber \\
&+\rho_{1q} \Big[ - G_+^*(\tau) X_{+,j}^{qn}(-\tau) - i G_-^*(\tau) X_{-,j}^{qn}(-\tau) \Big] \delta_{mi} \nonumber \\
&+\rho_{iq} \Big[ - G_+^*(\tau) X_{+,j}^{qn}(-\tau) + i G_-^*(\tau) X_{-,j}^{qn}(-\tau) \Big] \delta_{m1} \Big).
\label{eq:memory_kernel}
\end{align}
Here, we defined system correlations $X_{+,i}(\tau) := \left( \sigma_{1i}(\tau) + \sigma_{i1}(\tau) \right)$, $X_{-,i}(\tau) := i\left( \sigma_{i1}(\tau) - \sigma_{1i}(\tau) \right)$ and denoted $\bra{m} X_{\pm,i}(-\tau) \ket{n} = X_{\pm,i}^{mn}(-\tau)$ for a shorter notation. The phonon correlation function is given by
\begin{align}
&\phi(\tau) = \int \! \mathrm{d}^3 {k} \ \dfrac{g_{\bm{k}}^2}{\omega_k^2} \left[ \coth \left( \dfrac{\hbar \omega_k}{2 k_B T} \right) \cos(\omega_k \tau) - i \sin(\omega_k \tau) \right],
\end{align}
with $G_+(\tau) := \cosh \left[ \phi(\tau) \right] -1$, $G_-(\tau) := \sinh \left[ \phi(\tau) \right]$ and $\bar{\Omega} = \Omega \exp \left[ - \phi(0)/2 \right]$.
This approach offers a high grade of numerical performance and stability, allowing for long simulation times and up to moderate and strong electron-phonon coupling strengths~\cite{Glaessl2011, Chow2013, Reiter2014, Reiter2019}.

\begin{figure}[t]
\centering
\includegraphics[width=\linewidth]{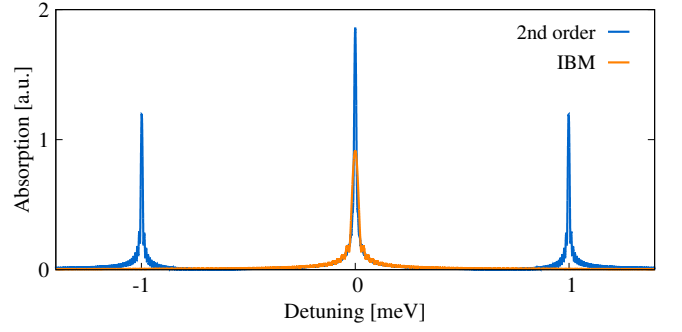}
\caption{Linear absorption spectrum of the considered \textit{V}-type emitter, calculated in the Heisenberg picture up to second order (blue line). The corresponding absorption spectrum of the analytically solvable \textit{Independent Boson Model} is shown in orange.}
\label{fig:spectrum}
\end{figure}

\begin{figure*}[t]
\centering
\includegraphics[width=\textwidth]{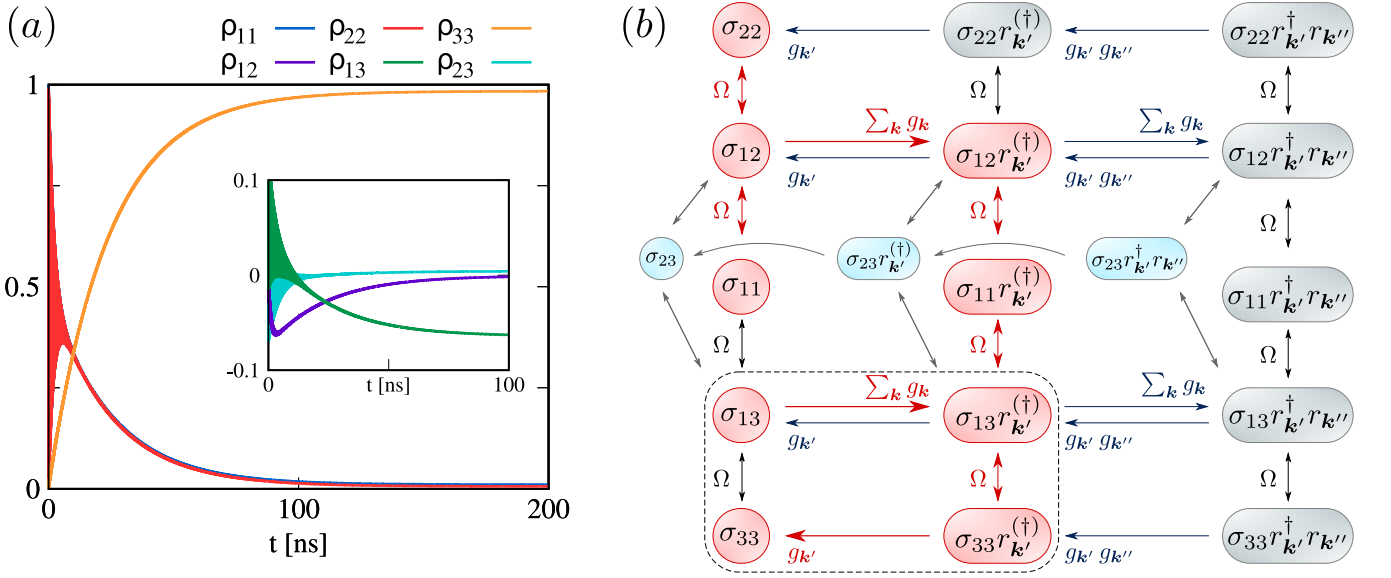}
\caption{(a) Occupation dynamics of a single $V$-type emitter calculated using the polaron master equation, exhibiting complete population inversion in the red-shifted excited state $\ket{3}$. The inset shows the corresponding coherence dynamics. (b) Diagrammatric representation of the Heisenberg equations of motion for a \textit{V}-emitter up to second order in phonon contributions. Red colored shapes and arrows represent the crucial interactions resulting in the observed population inversion. Turquois shapes illustrate the role of coherence $\sigma_{23}$ and phonon-assisted amplitudes.}
\label{fig:dynamics}
\end{figure*}

However, to unravel the underlying physical mechanisms, we also calculate the dynamics in the Heisenberg picture up to second order in the phonon contributions~\cite{Schilp1994, Krummheuer2002, Rossi2002, Foerstner2003, Foerstner2003pssb, su2013collective, Reiter2019}. To demonstrate the equivalence of the two descriptions in the considered parameter regime, we show a comparison of the dynamics resulting from both approaches in Appendix~\ref{app:vs}.
The Heisenberg equations of motion for the bare electronic contributions are given by $-i\hbar \partial_t \sigma_{mn}(t)=[H,\sigma_{mn} (t)]$, yielding
\begin{align}
&\partial_t \braket{\sigma_{mn}} = i \Big[ \Delta_2 \left( \braket{\sigma_{2n}} \delta_{m2} - \braket{\sigma_{m2}} \delta_{n2} \right) \nonumber \\
&+ \Delta_3 \left( \braket{\sigma_{3n}} \delta_{m3} - \braket{\sigma_{m3}} \delta_{n3} \right) \nonumber \\
&+ \Omega \Big( \braket{\sigma_{1n}} \delta_{m2} + \braket{\sigma_{2n}} \delta_{m1} - \braket{\sigma_{m2}} \delta_{n1} - \braket{\sigma_{m1}} \delta_{n2} \nonumber \\
& + \braket{\sigma_{1n}} \delta_{m3} + \braket{\sigma_{3n}} \delta_{m1} - \braket{\sigma_{m3}} \delta_{n1} - \braket{\sigma_{m1}} \delta_{n3} \Big) \nonumber \\
&+ \int \mathrm{d} \bm{k} \  g_{\bm{k}} \Big( \braket{\sigma_{2n} r_{\bm{k}}^\dagger} \delta_{m2} + \braket{\sigma_{3n} r_{\bm{k}}^\dagger} \delta_{m3} - \braket{\sigma_{m2} r_{\bm{k}}^\dagger} \delta_{n2} \nonumber \\
& - \braket{\sigma_{m3} r_{\bm{k}}^\dagger} \delta_{n3} + \braket{\sigma_{2n} r_{\bm{k}}} \delta_{m2} + \braket{\sigma_{3n} r_{\bm{k}}} \delta_{m3} \nonumber \\
& - \braket{\sigma_{m2} r_{\bm{k}}} \delta_{n2} - \braket{\sigma_{m3} r_{\bm{k}}} \delta_{n3} \Big) \Big],
\label{eq:heisenberg}
\end{align}
where $m,n=\{1,2,3\}$ number the different levels and the time argument of all operators is not written explicitly.
The equations of motion for the phonon-assisted transitions ($m \neq n$) and occupations ($m=n$) $\braket{\sigma_{mn} r_{\bm{k}}^{(\dagger)}}$ are given in Appendix~\ref{app:eqs_motion}. To close the system of equations, a second-order Born factorization is employed in the limit where a correlation expansion reproduces the full \textit{Independent Boson Model} (IBM, cf. Fig.~\ref{fig:spectrum}). The mean phonon number is assumed as a thermal Bose distribution~\cite{Breuer2002}, $\braket{r_{\bm{k}}^\dagger r_{\bm{k}}} \approx \{ \exp[\hbar \omega_k /(k_B T)]-1 \}^{-1}$, with $k_B$ the Boltzmann constant.
To illustrate the involved oscillator strengths, Fig.~\ref{fig:spectrum} shows the resulting linear absorption spectrum $\alpha (\omega) \sim \mathrm{Im} [\sigma_{12}(\omega)] + \mathrm{Im} [\sigma_{13} (\omega)]$~\cite{Mukamel1999, Malic2013, Carmele2019} (blue line), featuring two side peaks resulting from absorption processes of the two phonon-dressed states. The center peak corresponds to the incoming laser frequency. For comparison, the absorption spectrum of the exactly solvable IBM~\cite{Mahan2000, Krummheuer2002, Foerstner2002} is shown at the same temperature and using the same material parameters for a two level system (orange line), featuring a single absorption peak of the laser frequency.


\section{Complete population inversion by non-Markovian reservoir interaction}

First, we discuss the case of a single quantum dot. Initially, the electron is localized in the ground state $\ket{1}$. We choose the laser frequency in resonance with the transition between ground state $\ket{1}$ and excited state $\ket{2}$, i.e. $\Delta_2=0$ and therefore $\delta \epsilon = \hbar \Delta_3 \neq 0$, corresponding to a negative energy detuning ($\omega_3 < \omega_L$) of the order of acoustic phonon frequencies.
Fig.~\ref{fig:dynamics}(a) shows the emitter occupation dynamics $\sigma_{11}$, $\sigma_{22}$ and $\sigma_{33}$ after the switch-on of the laser field with corresponding Rabi frequency $\Omega$. In the course of time, the system achieves a complete population inversion: The detuned state $\ket{3}$ reaches full occupation $\braket{\sigma_{33}} \rightarrow 1$ in the steady state limit, while the populations of the resonantly driven levels $\ket{1}$ and $\ket{2}$ decline to zero while exhibiting oscillations. The subsequent analysis of this behavior is focused on the Heisenberg description due to its more intuitive nature.

To backtrace the interaction processes leading to the population inversion, we refer to Fig.~\ref{fig:dynamics}(b), illustrating the interactions between coherences, occupations and their phonon-assisted amplitudes in correspondence to their occurrence in the Heisenberg equations of motion [Eqs.~\eqref{eq:heisenberg}, \eqref{eq:motion2} and \eqref{eq:motion3}]: First, note that a direct optical transition from the resonantly driven excited state $\ket{2}$ to the detuned excited state $\ket{3}$ is forbidden. Hence the structured reservoir must facilitate the population transfer to level $\ket{3}$.
Red-colored shapes and arrows in Fig.~\ref{fig:dynamics}(b) indicate the key interactions of occupations and coherences responsible for the excitation transfer to level $\ket{3}$. In consequence of the continuous resonant driving of the transition between levels $\ket{1}$ and $\ket{2}$, energy is dissipated to the environment via the decay of the coherently driven transition amplitude $\sigma_{12}$: 
Enabled by the non-Markovian nature of the environment which allows for excitation backflow from the reservoir to the system, phonon-assisted transitions $\sigma_{ij}r_{\bm{k}}^{(\dagger)}$ form a gateway from the optically driven amplitude $\sigma_{12}$ to the detuned excited state $\sigma_{33}$, driven by the laser field $\Omega$ [cf. Fig.~\ref{fig:dynamics}(b)].
The detailed path is as follows: The excitation is transferred via phonon-assisted transitions from $\sigma_{12}r_{\bm{k}}^{(\dagger)}$ via $\sigma_{11}r_{\bm{k}}^{(\dagger)}$, $\sigma_{13}r_{\bm{k}}^{(\dagger)}$ and $\sigma_{33}r_{\bm{k}}^{(\dagger)}$ to the detuned excited state $\sigma_{33}$.
Once the energy is transferred to $\sigma_{33}$, laser field driven transitions back to the ground state could take place via $\sigma_{13}$, but dissipation to the phonon mode continuum occurs on a much faster time scale: Caused by non-Markovian system-reservoir interactions via $\sigma_{13}r_{\bm{k}}^{(\dagger)}$ and $\sigma_{33}r_{\bm{k}}^{(\dagger)}$, the excitation becomes dynamically trapped in the detuned state $\sigma_{33}$ [dashed frame in Fig.~\ref{fig:dynamics}(b)]. As the excitation in the two resonantly driven states is permanently lost by dissipation to the third level, their population uniformly decays towards zero.
The coherence $\sigma_{23}$ is an additional suspect of carrying out the transfer, since it constitutes the shortest connection between the two excited states via the coherences $\sigma_{12}$ and $\sigma_{13}$ [cf. Fig.~\ref{fig:dynamics}(b)]. However, switching off all contributions from $\sigma_{23}$ and $\sigma_{23} r_{\bm{k}^\prime}^{(\dagger)}$ still results in the full inversion exhibited in Fig.~\ref{fig:dynamics}(a). The corresponding dynamics of system coherences $\sigma_{12}$, $\sigma_{13}$ and $\sigma_{23}$ are shown in the inset. While $\sigma_{12}$ and $\sigma_{23}$ decay to zero over time, $\sigma_{13}$ takes on a finite value, as excitation leaking from the detuned level is constantly dissipated to the reservoir, preventing backflow to the ground state and allowing for the formation of a nonequilibrium steady state, which is not achievable in a Lindblad-based decoherence process.

\begin{figure}[t]
\centering
\includegraphics[width=0.8\linewidth]{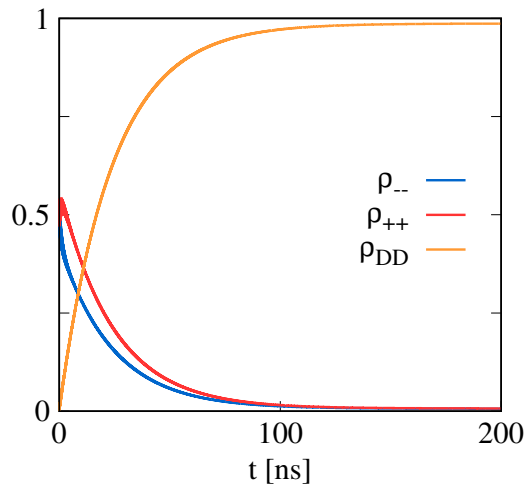}
\caption{Occupation dynamics of the density matrix eigenstates $\ket{-}$, $\ket{+}$ and $\ket{D}$, corresponding to Fig.~\ref{fig:dynamics}(a).}
\label{fig:eigenstates}
\end{figure}

Lastly, the emergence of the coherent population trapping can be viewed from a different perspective, as it is highly dependent on the energy detuning $\delta \epsilon$. In the absence of phonons, the system Hamiltonian reads in matrix form
\begin{equation}
H = \Omega
\begin{pmatrix}
  0 & 1 & 1 \\
  1 & 0 & 0 \\
  1 & 0 & \dfrac{\delta \epsilon}{\Omega}
 \end{pmatrix}.
\end{equation}
For the case of zero detuning, $\delta \epsilon = 0.0\,$meV, the corresponding eigenvalues can be calculated analytically,
\begin{equation}
\lambda_- = -\sqrt{2} \Omega,
\quad
\lambda_+ = \sqrt{2} \Omega,
\quad
\lambda_D = 0.
\end{equation}
Hence in the absence of phonons and at zero energy detuning between the two excited states, the eigenstates of the \textit{V}-emitter feature a dark state.
The corresponding eigenstates read in terms of the original basis states
\begin{align}
\ket{-} &= -\sqrt{2} \ket{1} + \ket{2} + \ket{3}, \nonumber \\
\ket{+} &= \sqrt{2} \ket{1} + \ket{2} + \ket{3}, \nonumber \\
\ket{D} &= \ket{2} - \ket{3}.
\end{align}
Fig.~\ref{fig:eigenstates} shows the occupation dynamics of the system density matrix eigenstates $\rho_{--}$, $\rho_{++}$ and $\rho_{DD}$, corresponding to the occupation dynamics shown in Fig.~\ref{fig:dynamics}(a). When considering an energy detuning $\delta \epsilon < 0$ together with the phonon reservoir, the former dark eigenstate $\ket{D}$ starts to interact weakly with the laser field and the reservoir and becomes completely occupied in a non-reciprocal phonon-mediated Stokes process (orange line). Analogous mechanisms have been shown to occur in exciton and biexciton systems under pulsed excitations in previous studies~\cite{Ardelt2014, Glaessl2011, Glaessl2013, Barth2016fast, Barth2016path}.


\section{Unidirectional quantum transport in a quantum dot chain}

\begin{figure*}[t]
\centering
\includegraphics[width=\linewidth]{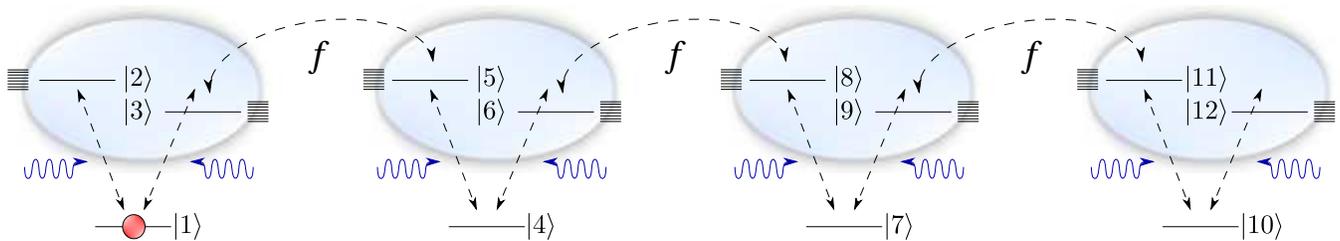}
\caption{A chain of \textit{V}-type emitters coupled by Dexter electron transfer with amplitude $f$, featuring a single interdot carrier transition due to electronic wave function overlap.}
\label{fig:vchain_dexter_simple}
\end{figure*}

To demonstrate the robustness of the population inversion and to provide an example of possible applications for quantum dot based coherent transport devices, we arrange multiple quantum dots in a chain-like, closely spaced configuration (cf. Fig.~\ref{fig:vchain_dexter_simple}).
As an example of an enabling resonance energy transfer mechanism between the quantum dots, we consider generic Dexter-type bidirectional \textit{electron transfer}~\cite{Dexter1953}, which has been shown to occur in self-assembled InAs-based quantum dot chains~\cite{Brusaferri1996, Sanguinetti2000, Tarasov2000, Mazur2002, Mazur2005, Mazur2009, Bhattacharyya2012}. It describes a Coulomb-induced direct and spin-preserving exchange of electrons between semiconductor nanostructures, thus requiring electronic wave function overlap between donor and acceptor levels~\cite{Dexter1953,Specht2015}.

For the most simple realization, we consider the case where only two levels of adjacent quantum dots feature an electronic wave function overlap, namely between the detuned excited state of the first and the resonantly driven excited state of the second emitter of each pair. In Appendix~\ref{app:couplings} it is shown that the transport still occurs even if all excited states of neighboring quantum dots are Dexter-coupled.
Moreover, the single excitation case is assumed with a single electron initially located at the ground state of the first emitter. Physical realizations of this setup, shown schematically in Fig.~\ref{fig:vchain_dexter_simple}, can be achieved e.g. by utilizing empty intersubband transitions in the conduction bands for the required \textit{V}-type transitions and a doping of the first quantum dot.
The corresponding interdot coupling Hamiltonian takes the form
\begin{equation}
H_{D} = \hbar f \sum_{l=0}^{N-2} \left( \sigma_{(3+3l)(5+3l)} + \mathrm{H.c.} \right),
\end{equation}
with a Dexter coupling amplitude $f$. Moreover, we choose separate phonon reservoirs with coupling elements $g_{k_{l}}$ acting independently on each emitter. The resulting polaron master equation for a chain of $N$ emitters reads
\begin{align}
&\partial_t \rho_S (t) = - \dfrac{i}{\hbar} \left[ H_{p,0}, \rho_S (t) \right]
- \dfrac{\bar{\Omega}^2}{\hbar^2} \sum_{l,{l^\prime}=0}^{N-1}  \sum_{i,j=2,3} \nonumber \\
&\times \int_0^t d\tau \sum_{s=\pm} \left( G_s(\tau)  \left[ X_{s,j}^{[l]},{X_{s,i}^{[{l^\prime}]}}^\prime \rho_S \right] + \mathrm{H.c.}\right)
\end{align}
with $X_{+,i}^{[l]} (\tau)= \left( \sigma_{(1+3l)(i+3l)} (\tau)+ \sigma_{(i+3l)(1+3l)}(\tau) \right)$, $X_{-,i}^{[l]}(\tau)= i\left( \sigma_{(i+3l)(1+3l)}(\tau) - \sigma_{(1+3l)(i+3l)}(\tau) \right)$, and $H_{p,0}$ now including the interdot coupling Hamiltonian $H_D$.

\begin{figure}[t]
\centering
\includegraphics[width=\linewidth]{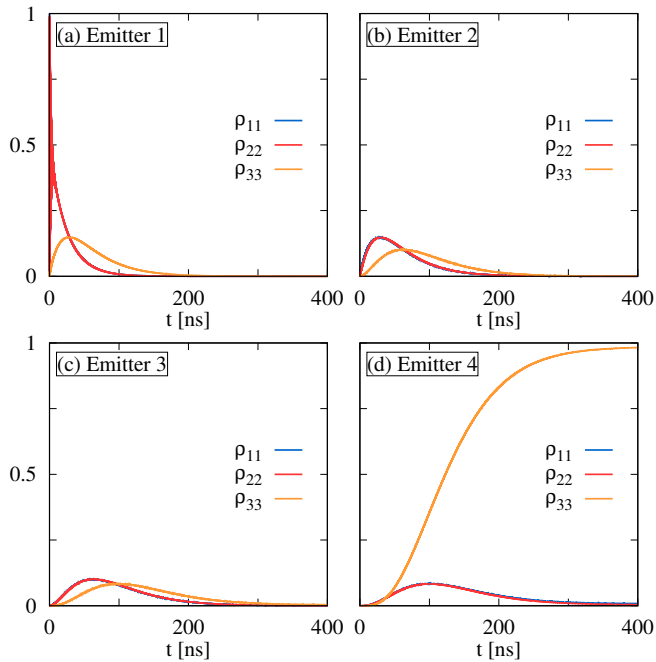}
\caption{Occupation dynamics for a chain of $N=4$ \textit{V}-type quantum dots coupled by Dexter-type electron transfer with a single interdot transition. Parameters are set to $f=0.1\,\mathrm{ps}^{-1}$ and $T=4\,$K, $\delta \epsilon=-1.0\,$meV, $\Omega=0.1\,\mathrm{ps}^{-1}$ as in the single emitter calculations.}
\label{fig:dexter1}
\end{figure}

Fig.~\ref{fig:dexter1} shows the resulting population dynamics for $N=4$ emitters and the Dexter coupling amplitude set to $f=0.1\,\mathrm{ps}^{-1}$. Driven by the non-reciprocal, phonon-assisted energy transfer to the detuned excited state in each emitter, the excitation is transferred to the detuned state of the last emitter in the chain. This behavior resembles \textit{perfect unidirectional electron transport} and is enabled by the population inversion effect rather than the interdot coupling mechanism itself:
At early times, the first emitter population in level $\ket{3}$ rises at the same rate as in the single emitter case [cf. Fig.~\ref{fig:dynamics}(b)]. Above a threshold occupation, Dexter interactions with the adjacent emitter start to dominate, enabling electron transfer and resulting in an asymptotic decline of $\ket{3}$ towards zero, while the resonantly driven levels $\ket{4}$ and $\ket{5}$ of the second emitter become evenly occupied.
Once excitation is transferred, the inversion process commences in the second emitter, resulting in a quick decline of population in levels $\ket{4}$ and $\ket{5}$ and in turn rising occupation in $\ket{6}$. Again, carrier transfer to the third quantum emitter is enabled above a threshold occupation. In consequence, all excitation is eventually transferred to the detuned excited state of the last emitter.

\begin{figure}[t]
\centering
\includegraphics[width=\linewidth]{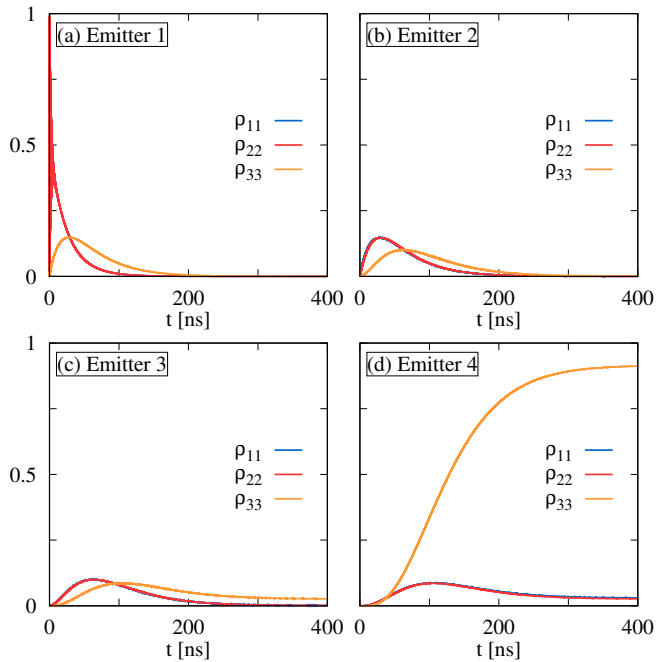}
\caption{Polaron master equation dynamics for a chain of $N=4$ \textit{V}-type emitters including radiative decay with $\gamma_r=0.1\,\mathrm{ns}^{-1}$. The remaining parameters are set equal to the calculations shown in Fig.~\ref{fig:dexter1}.}
\label{fig:vchain_radiative}
\end{figure}

Due to the robust nature of the inversion mechanism, the transport withstands perturbations such as radiative decay. We account for radiative dissipation from the excited states to the ground state by including a Lindblad dissipator of the form
\begin{align}
\mathcal{L}_{rad} \rho =& \gamma_r \sum_{l=1}^{N-1} \sum_{i=2,3} \big( 2 \sigma_{(1+3l)(i+3l)} \rho \sigma_{(i+3l)(1+3l)} \nonumber \\
& - \sigma_{(i+3l)(i+3l)} \rho - \rho \sigma_{(i+3l)(i+3l)} \big)
\end{align}
to the equations of motion for the density matrix, with $\gamma_r$ denoting the radiative decay rate. Fig.~\ref{fig:vchain_radiative} shows the quantum dot chain occupation dynamics corresponding to Fig.~\ref{fig:dexter1}, but including radiative decay at a rate $\gamma_r=0.1\,\mathrm{ns}^{-1}$.
In this case, the steady state occupation in the detuned excited state of the last emitter is decreased. However, the excitation transfer itself is not suppressed by the decay. As a result of continuous driving, the population is still transmitted to the last site of the emitter array, albeit with a decreased level of inversion depending on the radiative decay rate to the ground state.

Nondiagonal electron-phonon coupling between the excited states of the emitters imposes an additional perturbation that may occur in realistic scenarios. We show in Appendix~\ref{app:extensions} that the inclusion of intraband coupling slows merely down the process of population inversion, but does not change the qualitative behavior of the occupation dynamics.
Lastly, to demonstrate the emergence of undirectional quantum transport \textit{regardless} of the specific realization of bidirectional interdot coupling in a specific semiconductor setup, in Appendix~\ref{app:couplings} we additionally consider (i) Dexter coupling where all excited states of adjacent emitters feature electronic wave function overlap (cf. Figs.~\ref{fig:dexter2_de01}, \ref{fig:dexter2_de00}) and (ii) Förster coupling between the quantum dots~\cite{Foerster1948} (cf. Fig.~\ref{fig:foerster}), both exhibiting comparable unidirectional transport dynamics. In the latter case, dipole-dipole interactions induce \textit{excitation energy transfer}, which has been shown to occur in materials containing arrays of quantum dots~\cite{Crooker2002, Clapp2006, Rozbicki2008, Machnikowski2009}.
Exploiting the population inversion towards the red-detuned excited states, we have created unidirectional spatio-temporal transport of both excitation and carriers from one end of the quantum dot chain to the other, depending on the employed interdot coupling. Herewith, we have provided an example in which unidirectional, non-reciprocal quantum transport results without an externally applied bias, but from the interplay between coherence and incoherence in the system itself.


\section{Concluding discussions}

We have presented a highly nonequilibrium steady state preparation with complete population inversion in a continuously driven \textit{V}-type emitter, enabled by non-Markovian system-reservoir interactions and an energy detuning between the two excited states. The investigated system can be realized by InAs/GaAs quantum dots.
Notably, full inversion can be achieved for a wide range of coupling and driving parameters, underlining the robustness of the energy transfer. Moreover, the mechanism is found to be robust against perturbations such as phonon-induced intraband couplings, which inevitably arise in the case of a small energy detuning between the excited states. While the attainment of inversion is slightly slowed down in the presence of intraband phonon coupling, there are no qualitative changes in the overall dynamics.

As an example application, we have demonstrated the emergence of unidirectional excitation and carrier transport in an array of quantum dots, enabled by the population inversion effect, and demonstrated its robustness against the specific realization of bidirectional interdot coupling. The presence of radiative decay from the upper levels yields a decreased level of inversion in the last site depending on the decay rate, however, the transport mechanism itself is not disturbed by the damping as long as the radiative decay is not the dominant process in the system.
In summary, the presented non-Markovian enabled population inversion can be achieved and maintained under realistic experimental conditions. This mechanism gives rise to a wide range of possible applications utilizing electronic transport in quantum optical devices, corresponding to recent propositions of photonic unidirectional quantum transport in similar setups~\cite{Yuan2015,Lin2014,Peano2015,Peano2016,Bandres2016,Posha2019,Mahmo2020}. The predicted effect may also be relevant for biological systems, where excitation transport plays an important role in the context of light harvesting phenomena~\cite{Cupellini2018}.


\appendix

\begin{figure}[b]
\centering
\includegraphics[width=\linewidth]{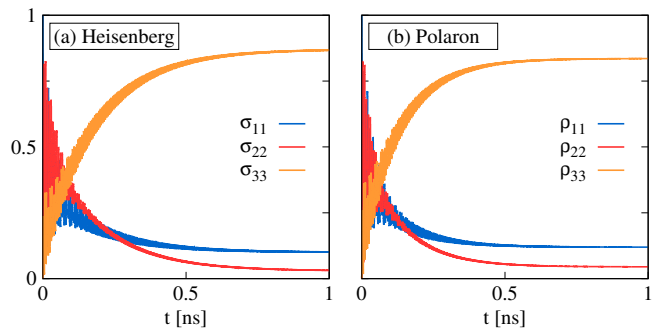}
\caption{Occupation dynamics of a single $V$-type emitter calculated (a) using the Heisenberg approach and (b) using the polaron master equation.}
\label{fig:versus}
\end{figure}

\section{Comparison between Heisenberg and Polaron Description} \label{app:vs}

In this work, we employed two different descriptions of the investigated \textit{V}-type system: The Heisenberg equation of motion approach up to second order in phonon contributions and the second order perturbative polaron master equation. The latter was used for numerical calculations, as it allows for longer simulation times at increased performance and stability. The Heisenberg approach, on the other hand, requires a very fine time discretization and breaks down after a few million time steps, thus limiting the accessible time scales.
To show the equivalence of the two descriptions, we employ a parameter set where both numerical implementations run stable and a steady state of the system is reached within the accessible simulation times. Fig.~\ref{fig:versus} shows occupation dynamics at $T=4\,$K, $\delta \epsilon=-1.0\,$meV, $\Omega=0.5\,\mathrm{ps}^{-1}$ and a rescaled coupling element $\tilde{g}_k=2.5g_k$, calculated using (a) the Heisenberg approach and (b) the polaron master equation.
Aside from a small phase shift introduced by the polaron transformation, which results in slightly different steady state occupations, the two descriptions yield very similar results. Although full inversion is not achieved in the accessible times, the qualitative and quantitative behavior is in very good agreement between the two descriptions, underlining their equivalence in the investigated parameter regime.


\section{Intraband Phonon Coupling} \label{app:extensions}

\begin{figure}[t]
\centering
\includegraphics[width=0.6\linewidth]{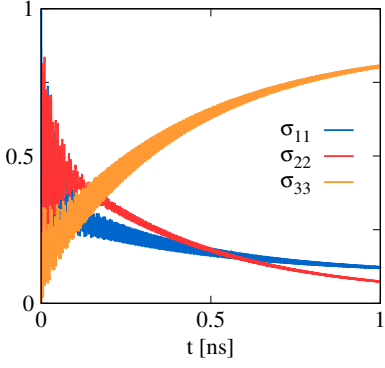}
\caption{Heisenberg picture dynamics for a single \textit{V}-type emitter including intraband coupling $g_{\bm{k}}^{intra}=0.1g_{\bm{k}}$. Parameters are set identical to the calculations shown in Fig.~\ref{fig:versus}.}
\label{fig:heisenberg_intra}
\end{figure}

Intraband phonon couplings between the two excited states of the \textit{V}-type emitter so far have been neglected in the electron-phonon interaction Hamiltonian (Eq.~\eqref{eq:H_elph}). However, in any realistic model system featuring excited states with low energy detunings, such couplings may arise. To validate the approximation of neglecting them from our model, we calculate the occupation dynamics once more using an extended interaction Hamiltonian of the form
\begin{align}
H_{el,ph} = &\hbar \! \int \! \mathrm{d}^3 k \  \big[ g_{\bm{k}} \left( \sigma_{22} + \sigma_{33} \right) \nonumber \\
& + g_{\bm{k}}^{intra} \left( \sigma_{23} + \sigma_{32} \right) \big] \left( r_{\bm{k}}^\dagger + r_{\bm{k}} \right),
\end{align}
explicitly including real-valued intraband couplings $g_{\bm{k}}^{intra}$ between the excited states.
Fig.~\ref{fig:heisenberg_intra} shows Heisenberg picture dynamics for a single emitter with unchanged parameters with respect to the dynamics shown in Fig.~\ref{fig:versus}(a), but including intraband phonon couplings. The intraband coupling strength has been set to a tenth of the interband coupling, $g_{\bm{k}}^{intra}=0.1g_{\bm{k}}$. Albeit the process of inversion is slightly slowed down, the dynamical behavior does not change qualitatively and hence justifies the approximation of disregarding intraband couplings in our case study.

\section{Heisenberg Equations of Motion} \label{app:eqs_motion}

The remaining Heisenberg equations of motion up to second order in phonon contributions for a single \textit{V}-type emitter read
\begin{align}
&\partial_t \braket{\sigma_{mn} r_{\bm{k}}^\dagger} \nonumber \\
= i \Big[& \Delta_2 \left( \braket{\sigma_{2n} r_{\bm{k}}^\dagger} \delta_{m2} - \braket{\sigma_{m2} r_{\bm{k}}^\dagger} \delta_{n2} \right) \nonumber \\
+ &\Delta_3 \left( \braket{\sigma_{3n} r_{\bm{k}}^\dagger} \delta_{m3} - \braket{\sigma_{m3} r_{\bm{k}}^\dagger} \delta_{n3} \right) \nonumber \\
+ &\Omega \Big( \braket{\sigma_{1n} r_{\bm{k}}^\dagger} \delta_{m2} + \braket{\sigma_{2n} r_{\bm{k}}^\dagger} \delta_{m1} - \braket{\sigma_{m2} r_{\bm{k}}^\dagger} \delta_{n1} \nonumber \\
- &\braket{\sigma_{m1} r_{\bm{k}}^\dagger} \delta_{n2} + \braket{\sigma_{1n} r_{\bm{k}}^\dagger} \delta_{m3} + \braket{\sigma_{3n} r_{\bm{k}}^\dagger} \delta_{m1} \nonumber \\
- &\braket{\sigma_{m3} r_{\bm{k}}^\dagger} \delta_{n1} - \braket{\sigma_{m1} r_{\bm{k}}^\dagger} \delta_{n3} \Big) 
+ \omega_k \braket{\sigma_{mn} r_{\bm{k}}^\dagger} \nonumber \\
+ &g_{\bm{k}} \left( \braket{r_{\bm{k}}^\dagger r_{\bm{k}}} +1 \right) \left( \braket{\sigma_{2n}} \delta_{m2} + \braket{\sigma_{3n}} \delta_{m3} \right) \nonumber \\
- &g_{\bm{k}} \braket{r_{\bm{k}}^\dagger r_{\bm{k}}} \left( \braket{\sigma_{m2}} \delta_{n2} + \braket{\sigma_{m3}} \delta_{n3} \right) \Big],
\label{eq:motion2}
\end{align}
\begin{align}
&\partial_t \braket{\sigma_{mn} r_{\bm{k}}} \nonumber \\
= i \Big[& \Delta_2 \left( \braket{\sigma_{2n} r_{\bm{k}}} \delta_{m2} - \braket{\sigma_{m2} r_{\bm{k}}} \delta_{n2} \right) \nonumber \\
+ &\Delta_3 \left( \braket{\sigma_{3n} r_{\bm{k}}} \delta_{m3} - \braket{\sigma_{m3} r_{\bm{k}}} \delta_{n3} \right) \nonumber \\
+ &\Omega \Big( \braket{\sigma_{1n} r_{\bm{k}}} \delta_{m2} + \braket{\sigma_{2n} r_{\bm{k}}} \delta_{m1} - \braket{\sigma_{m2} r_{\bm{k}}} \delta_{n1} \nonumber \\
- &\braket{\sigma_{m1} r_{\bm{k}}} \delta_{n2} + \braket{\sigma_{1n} r_{\bm{k}}} \delta_{m3} + \braket{\sigma_{3n} r_{\bm{k}}} \delta_{m1} \nonumber \\
- &\braket{\sigma_{m3} r_{\bm{k}}} \delta_{n1} - \braket{\sigma_{m1} r_{\bm{k}}} \delta_{n3} \Big) - \omega_k \braket{\sigma_{mn} r_{\bm{k}}} \nonumber \\
+ &g_{\bm{k}} \braket{r_{\bm{k}}^\dagger r_{\bm{k}}} \left( \braket{\sigma_{2n}} \delta_{m2} + \braket{\sigma_{3n}} \delta_{m3} \right) \nonumber \\
- &g_{\bm{k}} \left( \braket{r_{\bm{k}}^\dagger r_{\bm{k}}} +1 \right) \left( \braket{\sigma_{m2}} \delta_{n2} + \braket{\sigma_{m3}} \delta_{n3} \right) \Big].
\label{eq:motion3}
\end{align}


\section{Alternative Interdot Coupling Mechanisms} \label{app:couplings}

\subsection{Dexter Coupling}

\begin{figure}[b]
\centering
\includegraphics[width=\linewidth]{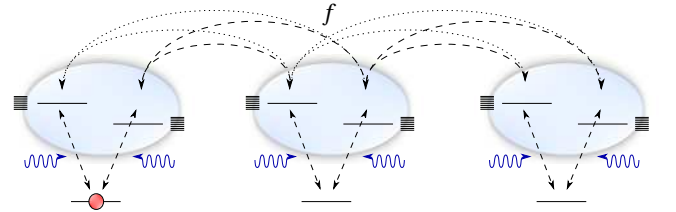}
\caption{A chain of \textit{V}-type emitters with interdot coupling enabled by Dexter-type carrier transfer interactions between all excited states of adjacent emitters with amplitude $f$. Initially, a single electron is located in the ground state of the first quantum dot.}
\label{fig:vchain_dexter}
\end{figure}

\begin{figure}[t]
\centering
\includegraphics[width=\linewidth]{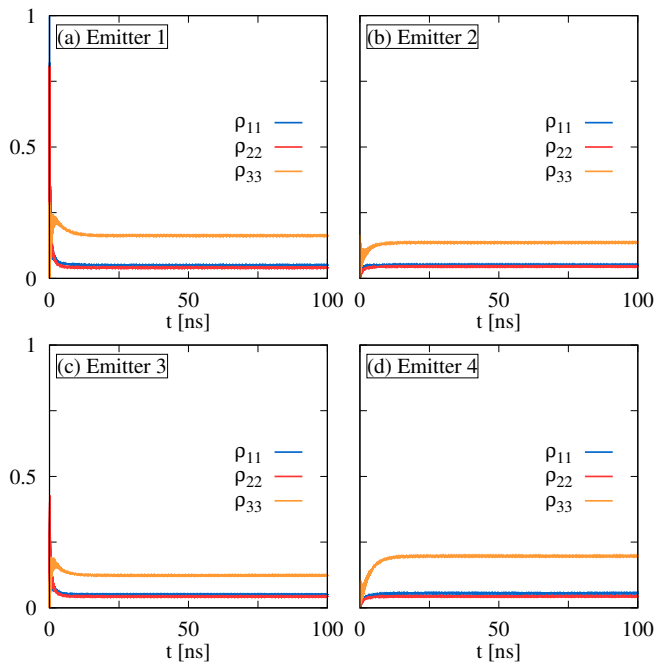}
\caption{Occupation dynamics for a chain of $N=4$ \textit{V}-type quantum dots with interdot coupling enabled by Dexter-type electron transfer processes, exhibiting a steady state current in the detuned excited state with the highest occupation located in the last emitter. Parameters are set to $f=0.1\,\mathrm{ps}^{-1}$, $\delta \epsilon=-1.0\,$meV, $\Omega=0.5\,\mathrm{ps}^{-1}$ and $T=4\,$K.}
\label{fig:dexter2_de01}
\end{figure}

\begin{figure}[t]
\centering
\includegraphics[width=\linewidth]{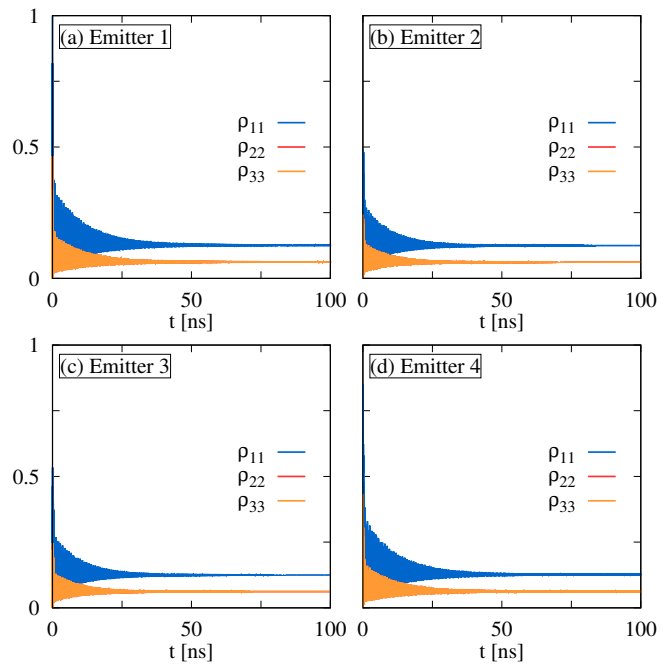}
\caption{Occupation dynamics corresponding to Fig.~\ref{fig:dexter2_de01}, but with the energy detuning set to $\delta \epsilon=0.0\,$meV. In this configuration, no steady state current is observed.}
\label{fig:dexter2_de00}
\end{figure}

To demonstrate the robustness of the unidirectional quantum transport phenomenon enabled by the population inversion effect, we consider two additional realizations of resonance energy transfer between the quantum dots~\cite{Specht2015}: Dexter-type direct \textit{electron transfer}~\cite{Dexter1953} and dipole-dipole interaction induced Förster \textit{excitation transfer}~\cite{Foerster1948}.
First, Dexter-type electron transfer is considered with electronic wave function overlaps between all excited states of adjacent emitters, as illustrated in Fig.~\ref{fig:vchain_dexter}. For simplicity, here we consider the single excitation case with a single electron initially located at the ground state of the first emitter. The corresponding coupling Hamiltonian reads
\begin{align}
H_{D} = &\hbar f  \sum_{l=0}^{N-2} ( \sigma_{(2+3l)(5+3l)} + \sigma_{(2+3l)(6+3l)} \nonumber \\
& + \sigma_{(3+3l)(5+3l)} + \sigma_{(3+3l)(6+3l)} + \mathrm{H.c.} ),
\end{align}
with a Dexter coupling amplitude $f$.
Fig.~\ref{fig:dexter2_de01} shows the resulting occupation dynamics in a chain of $N=4$ emitters. The Dexter coupling between all excited states of adjacent sites effectively yields a periodic steady state current in the chain. The detuned excited states of the individual chain sites carry a vast majority of the steady state occupation (orange lines), with the highest occupation located in the last quantum dot. Hence, we observe unidirectional electron transfer, which is enabled by the previously described population inversion effect due to non-Markovian system-reservoir interactions.
To underline the crucial role of the population inversion effect for the interdot electron transfer, we also calculate the chain dynamics at \textit{zero energy detuning}, $\delta \epsilon = 0.0\,$meV, where no population inversion takes place. Fig.~\ref{fig:dexter2_de00} shows the corresponding occupation dynamics. Again, the Dexter coupling induces carrier transport, albeit at a much lower rate and \textit{without} exhibiting a steady state current: The highest occupations in all emitters are located in the respective ground states, with identical steady state occupations in all quantum dots (blue lines), while the excited states of emitters 2, 3 and 4 are evenly occupied with a population close to zero. This proves that the interdot Dexter coupling is not the enabling factor for the unidirectional electron transfer exhibited in Fig.~\ref{fig:dexter2_de01}, but rather the non-Markovian population inversion.

\subsection{Förster Coupling}

\begin{figure}[t]
\centering
\includegraphics[width=\linewidth]{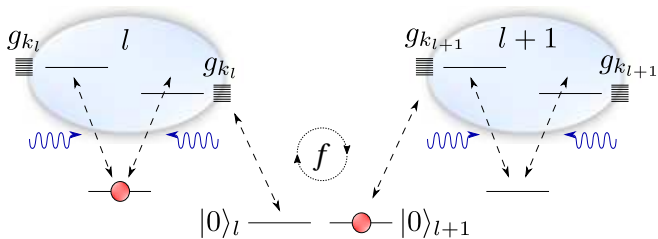}
\caption{Two identical \textit{V}-type emitters with an additional carrier reservoir state $\ket{0}$ to enable excitation transfer between systems. Interdot coupling is enabled via dipole-dipole interactions with an amplitude $f$, facilitating Förster transfer.}
\label{fig:vchain_dipole}
\end{figure}

Secondly, we consider bidirectional dipole-dipole induced Förster \textit{excitation energy transfer} between the quantum dots. Specifically, it was shown that optical near-field interactions, for instance caused by exciton-polariton coupling, can induce dipole-forbidden excitation energy transfer between quantum dots in close proximity~\cite{Kawazoe2002, Kobayashi2003, Sangu2003, Sangu2004, Sato2007, Nishibayashi2008}. The considered Förster-type dipole-forbidden level transitions are illustrated in Fig.~\ref{fig:vchain_dipole}. To facilitate the excitation transfer, an additional level $\ket{0}$ is added to each emitter, representing a wetting layer. The corresponding Förster coupling Hamiltonian reads
\begin{equation}
H_{F} = \hbar f \sum_{l=1}^{N-1} \left( \sigma_{03}^l \sigma_{20}^{l+1} + \mathrm{H.c.} \right),
\end{equation}
with $l$ denoting the chain sites and a coupling amplitude $f$.
Again, we initially assume a single electron in the ground state of the first emitter. Due to the high numerical cost of simulations including more than one excitation, calculations are performed for $N=2$ emitters. However, the qualitative dynamics are robust against increasing numbers of sites.

\begin{figure}[t]
\centering
\includegraphics[width=\linewidth]{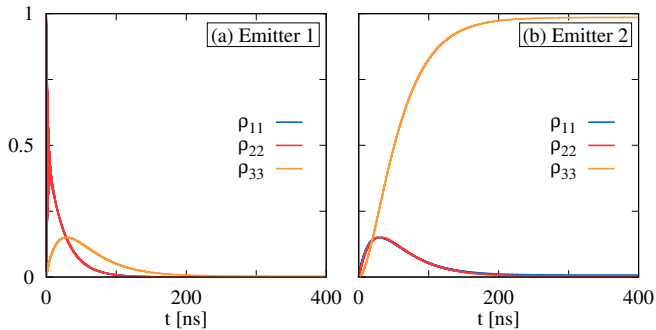}
\caption{Occupation dynamics for $N=2$ \textit{V}-type emitters coupled by Förster interactions, exhibiting unidirectional excitation transport from one emitter to the other. Parameters are set to $f=0.1\,\mathrm{ps}^{-1}$ and $T=4\,$K, $\delta \epsilon=-1.0\,$meV, $\Omega=0.1\,\mathrm{ps}^{-1}$ as in the single emitter case.}
\label{fig:foerster}
\end{figure}

Fig.~\ref{fig:foerster} shows the resulting population dynamics for $N=2$ emitters at $f=0.1\, \mathrm{ps}^{-1}$ with close resemblance to the dynamics exhibited in the single transition Dexter coupling case. Initially, the first emitter is prepared in the ground state while the remaining sites are initialized at levels $\ket{0}$. Driven by the non-reciprocal, phonon-assisted energy transfer to the detuned excited state in each emitter, the occupation is transferred to the detuned state of the second emitter.
At early times, the first emitter population in level $\ket{3}_1$ rises at the same rate as in the single emitter case. Above a threshold occupation, dipole-dipole interactions with the adjacent emitter start to dominate, enabling Förster resonance energy transfer and resulting in an asymptotic decline of $\ket{3}_1$ towards zero, while the resonantly driven levels $\ket{1}_2$ and $\ket{2}_2$ of the second emitter become evenly occupied.
Once excitation is transferred, the inversion process commences in the second emitter, resulting in a quick decline of population in levels $\ket{1}_2$ and $\ket{2}_2$ and in turn rising occupation in $\ket{3}_2$. In consequence, the excitation is swiftly divided between the detuned states $\ket{3}_1$ and $\ket{3}_2$ of the two emitters with the major share located in the last site. Eventually, all population is transferred to the detuned state of the last emitter.
With this, we have demonstrated the unique dependence of the unidirectional excitation transport on the population inversion effect exhibited in \textit{V}-type quantum dots, rather than the explicit mechanism of interdot coupling in a specific semiconductor setup.


\begin{acknowledgments}
O.K., A.K. and A.C. gratefully acknowledge support from the Deutsche Forschungsgemeinschaft (DFG) through SFB 910 project B1, project number 163436311. E.V.D. and J.M. acknowlegdge support from the Danish National Research Foundation, grant number DNRF147.
\end{acknowledgments}


\begin{thebibliography}{104}%
\makeatletter
\providecommand \@ifxundefined [1]{%
 \@ifx{#1\undefined}
}%
\providecommand \@ifnum [1]{%
 \ifnum #1\expandafter \@firstoftwo
 \else \expandafter \@secondoftwo
 \fi
}%
\providecommand \@ifx [1]{%
 \ifx #1\expandafter \@firstoftwo
 \else \expandafter \@secondoftwo
 \fi
}%
\providecommand \natexlab [1]{#1}%
\providecommand \enquote  [1]{``#1''}%
\providecommand \bibnamefont  [1]{#1}%
\providecommand \bibfnamefont [1]{#1}%
\providecommand \citenamefont [1]{#1}%
\providecommand \href@noop [0]{\@secondoftwo}%
\providecommand \href [0]{\begingroup \@sanitize@url \@href}%
\providecommand \@href[1]{\@@startlink{#1}\@@href}%
\providecommand \@@href[1]{\endgroup#1\@@endlink}%
\providecommand \@sanitize@url [0]{\catcode `\\12\catcode `\$12\catcode
  `\&12\catcode `\#12\catcode `\^12\catcode `\_12\catcode `\%12\relax}%
\providecommand \@@startlink[1]{}%
\providecommand \@@endlink[0]{}%
\providecommand \url  [0]{\begingroup\@sanitize@url \@url }%
\providecommand \@url [1]{\endgroup\@href {#1}{\urlprefix }}%
\providecommand \urlprefix  [0]{URL }%
\providecommand \Eprint [0]{\href }%
\providecommand \doibase [0]{http://dx.doi.org/}%
\providecommand \selectlanguage [0]{\@gobble}%
\providecommand \bibinfo  [0]{\@secondoftwo}%
\providecommand \bibfield  [0]{\@secondoftwo}%
\providecommand \translation [1]{[#1]}%
\providecommand \BibitemOpen [0]{}%
\providecommand \bibitemStop [0]{}%
\providecommand \bibitemNoStop [0]{.\EOS\space}%
\providecommand \EOS [0]{\spacefactor3000\relax}%
\providecommand \BibitemShut  [1]{\csname bibitem#1\endcsname}%
\let\auto@bib@innerbib\@empty
\bibitem [{\citenamefont {Liu}\ \emph {et~al.}(2011)\citenamefont {Liu},
  \citenamefont {Li}, \citenamefont {Huang}, \citenamefont {Li}, \citenamefont
  {Guo}, \citenamefont {Laine}, \citenamefont {Breuer},\ and\ \citenamefont
  {Piilo}}]{Liu2011}%
  \BibitemOpen
  \bibfield  {author} {\bibinfo {author} {\bibfnamefont {B.-H.}\ \bibnamefont
  {Liu}}, \bibinfo {author} {\bibfnamefont {L.}~\bibnamefont {Li}}, \bibinfo
  {author} {\bibfnamefont {Y.-F.}\ \bibnamefont {Huang}}, \bibinfo {author}
  {\bibfnamefont {C.-F.}\ \bibnamefont {Li}}, \bibinfo {author} {\bibfnamefont
  {G.-C.}\ \bibnamefont {Guo}}, \bibinfo {author} {\bibfnamefont {E.-M.}\
  \bibnamefont {Laine}}, \bibinfo {author} {\bibfnamefont {H.-P.}\ \bibnamefont
  {Breuer}}, \ and\ \bibinfo {author} {\bibfnamefont {J.}~\bibnamefont
  {Piilo}},\ }\href {\doibase 10.1038/nphys2085} {\bibfield  {journal}
  {\bibinfo  {journal} {Nature Physics}\ }\textbf {\bibinfo {volume} {7}},\
  \bibinfo {pages} {931} (\bibinfo {year} {2011})}\BibitemShut {NoStop}%
\bibitem [{\citenamefont {Liu}\ \emph {et~al.}(2013)\citenamefont {Liu},
  \citenamefont {Cao}, \citenamefont {Huang}, \citenamefont {Li}, \citenamefont
  {Guo}, \citenamefont {Laine}, \citenamefont {Breuer},\ and\ \citenamefont
  {Piilo}}]{Liu2013}%
  \BibitemOpen
  \bibfield  {author} {\bibinfo {author} {\bibfnamefont {B.-H.}\ \bibnamefont
  {Liu}}, \bibinfo {author} {\bibfnamefont {D.-Y.}\ \bibnamefont {Cao}},
  \bibinfo {author} {\bibfnamefont {Y.-F.}\ \bibnamefont {Huang}}, \bibinfo
  {author} {\bibfnamefont {C.-F.}\ \bibnamefont {Li}}, \bibinfo {author}
  {\bibfnamefont {G.-C.}\ \bibnamefont {Guo}}, \bibinfo {author} {\bibfnamefont
  {E.-M.}\ \bibnamefont {Laine}}, \bibinfo {author} {\bibfnamefont {H.-P.}\
  \bibnamefont {Breuer}}, \ and\ \bibinfo {author} {\bibfnamefont
  {J.}~\bibnamefont {Piilo}},\ }\href {\doibase 10.1038/srep01781} {\bibfield
  {journal} {\bibinfo  {journal} {Scientific Reports}\ }\textbf {\bibinfo
  {volume} {3}},\ \bibinfo {pages} {1781} (\bibinfo {year} {2013})}\BibitemShut
  {NoStop}%
\bibitem [{\citenamefont {Reich}\ \emph {et~al.}(2015)\citenamefont {Reich},
  \citenamefont {Katz},\ and\ \citenamefont {Koch}}]{Reich2015}%
  \BibitemOpen
  \bibfield  {author} {\bibinfo {author} {\bibfnamefont {D.~M.}\ \bibnamefont
  {Reich}}, \bibinfo {author} {\bibfnamefont {N.}~\bibnamefont {Katz}}, \ and\
  \bibinfo {author} {\bibfnamefont {C.~P.}\ \bibnamefont {Koch}},\ }\href
  {\doibase 10.1038/srep12430} {\bibfield  {journal} {\bibinfo  {journal}
  {Scientific Reports}\ }\textbf {\bibinfo {volume} {5}},\ \bibinfo {pages}
  {12430} (\bibinfo {year} {2015})}\BibitemShut {NoStop}%
\bibitem [{\citenamefont {Wolf}\ \emph {et~al.}(2008)\citenamefont {Wolf},
  \citenamefont {Eisert}, \citenamefont {Cubitt},\ and\ \citenamefont
  {Cirac}}]{Wolf2008}%
  \BibitemOpen
  \bibfield  {author} {\bibinfo {author} {\bibfnamefont {M.~M.}\ \bibnamefont
  {Wolf}}, \bibinfo {author} {\bibfnamefont {J.}~\bibnamefont {Eisert}},
  \bibinfo {author} {\bibfnamefont {T.~S.}\ \bibnamefont {Cubitt}}, \ and\
  \bibinfo {author} {\bibfnamefont {J.~I.}\ \bibnamefont {Cirac}},\ }\href
  {\doibase 10.1103/PhysRevLett.101.150402} {\bibfield  {journal} {\bibinfo
  {journal} {Phys. Rev. Lett.}\ }\textbf {\bibinfo {volume} {101}},\ \bibinfo
  {pages} {150402} (\bibinfo {year} {2008})}\BibitemShut {NoStop}%
\bibitem [{\citenamefont {Chan}\ \emph {et~al.}(2014)\citenamefont {Chan},
  \citenamefont {Lin}, \citenamefont {Yelin},\ and\ \citenamefont
  {Lukin}}]{Chan2014}%
  \BibitemOpen
  \bibfield  {author} {\bibinfo {author} {\bibfnamefont {C.-K.}\ \bibnamefont
  {Chan}}, \bibinfo {author} {\bibfnamefont {G.-D.}\ \bibnamefont {Lin}},
  \bibinfo {author} {\bibfnamefont {S.~F.}\ \bibnamefont {Yelin}}, \ and\
  \bibinfo {author} {\bibfnamefont {M.~D.}\ \bibnamefont {Lukin}},\ }\href
  {\doibase 10.1103/PhysRevA.89.042117} {\bibfield  {journal} {\bibinfo
  {journal} {Phys. Rev. A}\ }\textbf {\bibinfo {volume} {89}},\ \bibinfo
  {pages} {042117} (\bibinfo {year} {2014})}\BibitemShut {NoStop}%
\bibitem [{\citenamefont {Jing}\ and\ \citenamefont {Yu}(2010)}]{Jing2010}%
  \BibitemOpen
  \bibfield  {author} {\bibinfo {author} {\bibfnamefont {J.}~\bibnamefont
  {Jing}}\ and\ \bibinfo {author} {\bibfnamefont {T.}~\bibnamefont {Yu}},\
  }\href {\doibase 10.1103/PhysRevLett.105.240403} {\bibfield  {journal}
  {\bibinfo  {journal} {Phys. Rev. Lett.}\ }\textbf {\bibinfo {volume} {105}},\
  \bibinfo {pages} {240403} (\bibinfo {year} {2010})}\BibitemShut {NoStop}%
\bibitem [{\citenamefont {Kaer}\ \emph {et~al.}(2013)\citenamefont {Kaer},
  \citenamefont {Lodahl}, \citenamefont {Jauho},\ and\ \citenamefont
  {Mork}}]{Kaer2013}%
  \BibitemOpen
  \bibfield  {author} {\bibinfo {author} {\bibfnamefont {P.}~\bibnamefont
  {Kaer}}, \bibinfo {author} {\bibfnamefont {P.}~\bibnamefont {Lodahl}},
  \bibinfo {author} {\bibfnamefont {A.-P.}\ \bibnamefont {Jauho}}, \ and\
  \bibinfo {author} {\bibfnamefont {J.}~\bibnamefont {Mork}},\ }\href {\doibase
  10.1103/PhysRevB.87.081308} {\bibfield  {journal} {\bibinfo  {journal} {Phys.
  Rev. B}\ }\textbf {\bibinfo {volume} {87}},\ \bibinfo {pages} {081308}
  (\bibinfo {year} {2013})}\BibitemShut {NoStop}%
\bibitem [{\citenamefont {Kozlov}\ \emph {et~al.}(2006)\citenamefont {Kozlov},
  \citenamefont {Rostovtsev},\ and\ \citenamefont {Scully}}]{Kozlov2006}%
  \BibitemOpen
  \bibfield  {author} {\bibinfo {author} {\bibfnamefont {V.~V.}\ \bibnamefont
  {Kozlov}}, \bibinfo {author} {\bibfnamefont {Y.}~\bibnamefont {Rostovtsev}},
  \ and\ \bibinfo {author} {\bibfnamefont {M.~O.}\ \bibnamefont {Scully}},\
  }\href {\doibase 10.1103/PhysRevA.74.063829} {\bibfield  {journal} {\bibinfo
  {journal} {Phys. Rev. A}\ }\textbf {\bibinfo {volume} {74}},\ \bibinfo
  {pages} {063829} (\bibinfo {year} {2006})}\BibitemShut {NoStop}%
\bibitem [{\citenamefont {Stace}\ \emph {et~al.}(2005)\citenamefont {Stace},
  \citenamefont {Doherty},\ and\ \citenamefont {Barrett}}]{Stace2005}%
  \BibitemOpen
  \bibfield  {author} {\bibinfo {author} {\bibfnamefont {T.~M.}\ \bibnamefont
  {Stace}}, \bibinfo {author} {\bibfnamefont {A.~C.}\ \bibnamefont {Doherty}},
  \ and\ \bibinfo {author} {\bibfnamefont {S.~D.}\ \bibnamefont {Barrett}},\
  }\href {\doibase 10.1103/PhysRevLett.95.106801} {\bibfield  {journal}
  {\bibinfo  {journal} {Phys. Rev. Lett.}\ }\textbf {\bibinfo {volume} {95}},\
  \bibinfo {pages} {106801} (\bibinfo {year} {2005})}\BibitemShut {NoStop}%
\bibitem [{\citenamefont {Kaer}\ \emph {et~al.}(2010)\citenamefont {Kaer},
  \citenamefont {Nielsen}, \citenamefont {Lodahl}, \citenamefont {Jauho},\ and\
  \citenamefont {M\o{}rk}}]{Kaer2010}%
  \BibitemOpen
  \bibfield  {author} {\bibinfo {author} {\bibfnamefont {P.}~\bibnamefont
  {Kaer}}, \bibinfo {author} {\bibfnamefont {T.~R.}\ \bibnamefont {Nielsen}},
  \bibinfo {author} {\bibfnamefont {P.}~\bibnamefont {Lodahl}}, \bibinfo
  {author} {\bibfnamefont {A.-P.}\ \bibnamefont {Jauho}}, \ and\ \bibinfo
  {author} {\bibfnamefont {J.}~\bibnamefont {M\o{}rk}},\ }\href {\doibase
  10.1103/PhysRevLett.104.157401} {\bibfield  {journal} {\bibinfo  {journal}
  {Phys. Rev. Lett.}\ }\textbf {\bibinfo {volume} {104}},\ \bibinfo {pages}
  {157401} (\bibinfo {year} {2010})}\BibitemShut {NoStop}%
\bibitem [{\citenamefont {Weiler}\ \emph {et~al.}(2012)\citenamefont {Weiler},
  \citenamefont {Ulhaq}, \citenamefont {Ulrich}, \citenamefont {Richter},
  \citenamefont {Jetter}, \citenamefont {Michler}, \citenamefont {Roy},\ and\
  \citenamefont {Hughes}}]{Weiler2012}%
  \BibitemOpen
  \bibfield  {author} {\bibinfo {author} {\bibfnamefont {S.}~\bibnamefont
  {Weiler}}, \bibinfo {author} {\bibfnamefont {A.}~\bibnamefont {Ulhaq}},
  \bibinfo {author} {\bibfnamefont {S.~M.}\ \bibnamefont {Ulrich}}, \bibinfo
  {author} {\bibfnamefont {D.}~\bibnamefont {Richter}}, \bibinfo {author}
  {\bibfnamefont {M.}~\bibnamefont {Jetter}}, \bibinfo {author} {\bibfnamefont
  {P.}~\bibnamefont {Michler}}, \bibinfo {author} {\bibfnamefont
  {C.}~\bibnamefont {Roy}}, \ and\ \bibinfo {author} {\bibfnamefont
  {S.}~\bibnamefont {Hughes}},\ }\href {\doibase 10.1103/PhysRevB.86.241304}
  {\bibfield  {journal} {\bibinfo  {journal} {Phys. Rev. B}\ }\textbf {\bibinfo
  {volume} {86}},\ \bibinfo {pages} {241304} (\bibinfo {year}
  {2012})}\BibitemShut {NoStop}%
\bibitem [{\citenamefont {Denning}\ \emph {et~al.}(2020)\citenamefont
  {Denning}, \citenamefont {Iles-Smith}, \citenamefont {Gregersen},\ and\
  \citenamefont {Mork}}]{Denning2020}%
  \BibitemOpen
  \bibfield  {author} {\bibinfo {author} {\bibfnamefont {E.~V.}\ \bibnamefont
  {Denning}}, \bibinfo {author} {\bibfnamefont {J.}~\bibnamefont {Iles-Smith}},
  \bibinfo {author} {\bibfnamefont {N.}~\bibnamefont {Gregersen}}, \ and\
  \bibinfo {author} {\bibfnamefont {J.}~\bibnamefont {Mork}},\ }\href {\doibase
  10.1364/OME.380601} {\bibfield  {journal} {\bibinfo  {journal} {Opt. Mater.
  Express}\ }\textbf {\bibinfo {volume} {10}},\ \bibinfo {pages} {222}
  (\bibinfo {year} {2020})}\BibitemShut {NoStop}%
\bibitem [{\citenamefont {Stock}\ \emph {et~al.}(2011)\citenamefont {Stock},
  \citenamefont {Dachner}, \citenamefont {Warming}, \citenamefont {Schliwa},
  \citenamefont {Lochmann}, \citenamefont {Hoffmann}, \citenamefont {Toropov},
  \citenamefont {Bakarov}, \citenamefont {Derebezov}, \citenamefont {Richter},
  \citenamefont {Haisler}, \citenamefont {Knorr},\ and\ \citenamefont
  {Bimberg}}]{Stock2011}%
  \BibitemOpen
  \bibfield  {author} {\bibinfo {author} {\bibfnamefont {E.}~\bibnamefont
  {Stock}}, \bibinfo {author} {\bibfnamefont {M.-R.}\ \bibnamefont {Dachner}},
  \bibinfo {author} {\bibfnamefont {T.}~\bibnamefont {Warming}}, \bibinfo
  {author} {\bibfnamefont {A.}~\bibnamefont {Schliwa}}, \bibinfo {author}
  {\bibfnamefont {A.}~\bibnamefont {Lochmann}}, \bibinfo {author}
  {\bibfnamefont {A.}~\bibnamefont {Hoffmann}}, \bibinfo {author}
  {\bibfnamefont {A.~I.}\ \bibnamefont {Toropov}}, \bibinfo {author}
  {\bibfnamefont {A.~K.}\ \bibnamefont {Bakarov}}, \bibinfo {author}
  {\bibfnamefont {I.~A.}\ \bibnamefont {Derebezov}}, \bibinfo {author}
  {\bibfnamefont {M.}~\bibnamefont {Richter}}, \bibinfo {author} {\bibfnamefont
  {V.~A.}\ \bibnamefont {Haisler}}, \bibinfo {author} {\bibfnamefont
  {A.}~\bibnamefont {Knorr}}, \ and\ \bibinfo {author} {\bibfnamefont
  {D.}~\bibnamefont {Bimberg}},\ }\href {\doibase 10.1103/PhysRevB.83.041304}
  {\bibfield  {journal} {\bibinfo  {journal} {Phys. Rev. B}\ }\textbf {\bibinfo
  {volume} {83}},\ \bibinfo {pages} {041304} (\bibinfo {year}
  {2011})}\BibitemShut {NoStop}%
\bibitem [{\citenamefont {Hughes}\ \emph {et~al.}(2011)\citenamefont {Hughes},
  \citenamefont {Yao}, \citenamefont {Milde}, \citenamefont {Knorr},
  \citenamefont {Dalacu}, \citenamefont {Mnaymneh}, \citenamefont {Sazonova},
  \citenamefont {Poole}, \citenamefont {Aers}, \citenamefont {Lapointe},
  \citenamefont {Cheriton},\ and\ \citenamefont {Williams}}]{Hughes2011}%
  \BibitemOpen
  \bibfield  {author} {\bibinfo {author} {\bibfnamefont {S.}~\bibnamefont
  {Hughes}}, \bibinfo {author} {\bibfnamefont {P.}~\bibnamefont {Yao}},
  \bibinfo {author} {\bibfnamefont {F.}~\bibnamefont {Milde}}, \bibinfo
  {author} {\bibfnamefont {A.}~\bibnamefont {Knorr}}, \bibinfo {author}
  {\bibfnamefont {D.}~\bibnamefont {Dalacu}}, \bibinfo {author} {\bibfnamefont
  {K.}~\bibnamefont {Mnaymneh}}, \bibinfo {author} {\bibfnamefont
  {V.}~\bibnamefont {Sazonova}}, \bibinfo {author} {\bibfnamefont {P.~J.}\
  \bibnamefont {Poole}}, \bibinfo {author} {\bibfnamefont {G.~C.}\ \bibnamefont
  {Aers}}, \bibinfo {author} {\bibfnamefont {J.}~\bibnamefont {Lapointe}},
  \bibinfo {author} {\bibfnamefont {R.}~\bibnamefont {Cheriton}}, \ and\
  \bibinfo {author} {\bibfnamefont {R.~L.}\ \bibnamefont {Williams}},\ }\href
  {\doibase 10.1103/PhysRevB.83.165313} {\bibfield  {journal} {\bibinfo
  {journal} {Phys. Rev. B}\ }\textbf {\bibinfo {volume} {83}},\ \bibinfo
  {pages} {165313} (\bibinfo {year} {2011})}\BibitemShut {NoStop}%
\bibitem [{\citenamefont {Thoma}\ \emph {et~al.}(2016)\citenamefont {Thoma},
  \citenamefont {Schnauber}, \citenamefont {Gschrey}, \citenamefont {Seifried},
  \citenamefont {Wolters}, \citenamefont {Schulze}, \citenamefont
  {Strittmatter}, \citenamefont {Rodt}, \citenamefont {Carmele}, \citenamefont
  {Knorr}, \citenamefont {Heindel},\ and\ \citenamefont
  {Reitzenstein}}]{Thoma2016}%
  \BibitemOpen
  \bibfield  {author} {\bibinfo {author} {\bibfnamefont {A.}~\bibnamefont
  {Thoma}}, \bibinfo {author} {\bibfnamefont {P.}~\bibnamefont {Schnauber}},
  \bibinfo {author} {\bibfnamefont {M.}~\bibnamefont {Gschrey}}, \bibinfo
  {author} {\bibfnamefont {M.}~\bibnamefont {Seifried}}, \bibinfo {author}
  {\bibfnamefont {J.}~\bibnamefont {Wolters}}, \bibinfo {author} {\bibfnamefont
  {J.-H.}\ \bibnamefont {Schulze}}, \bibinfo {author} {\bibfnamefont
  {A.}~\bibnamefont {Strittmatter}}, \bibinfo {author} {\bibfnamefont
  {S.}~\bibnamefont {Rodt}}, \bibinfo {author} {\bibfnamefont {A.}~\bibnamefont
  {Carmele}}, \bibinfo {author} {\bibfnamefont {A.}~\bibnamefont {Knorr}},
  \bibinfo {author} {\bibfnamefont {T.}~\bibnamefont {Heindel}}, \ and\
  \bibinfo {author} {\bibfnamefont {S.}~\bibnamefont {Reitzenstein}},\ }\href
  {\doibase 10.1103/PhysRevLett.116.033601} {\bibfield  {journal} {\bibinfo
  {journal} {Phys. Rev. Lett.}\ }\textbf {\bibinfo {volume} {116}},\ \bibinfo
  {pages} {033601} (\bibinfo {year} {2016})}\BibitemShut {NoStop}%
\bibitem [{\citenamefont {Su}\ \emph {et~al.}(2013{\natexlab{a}})\citenamefont
  {Su}, \citenamefont {Bimberg}, \citenamefont {Knorr},\ and\ \citenamefont
  {Carmele}}]{Su2013}%
  \BibitemOpen
  \bibfield  {author} {\bibinfo {author} {\bibfnamefont {Y.}~\bibnamefont
  {Su}}, \bibinfo {author} {\bibfnamefont {D.}~\bibnamefont {Bimberg}},
  \bibinfo {author} {\bibfnamefont {A.}~\bibnamefont {Knorr}}, \ and\ \bibinfo
  {author} {\bibfnamefont {A.}~\bibnamefont {Carmele}},\ }\href {\doibase
  10.1103/PhysRevLett.110.113604} {\bibfield  {journal} {\bibinfo  {journal}
  {Phys. Rev. Lett.}\ }\textbf {\bibinfo {volume} {110}},\ \bibinfo {pages}
  {113604} (\bibinfo {year} {2013}{\natexlab{a}})}\BibitemShut {NoStop}%
\bibitem [{\citenamefont {Kabuss}\ \emph
  {et~al.}(2011{\natexlab{a}})\citenamefont {Kabuss}, \citenamefont {Carmele},
  \citenamefont {Richter},\ and\ \citenamefont {Knorr}}]{Kabuss2011}%
  \BibitemOpen
  \bibfield  {author} {\bibinfo {author} {\bibfnamefont {J.}~\bibnamefont
  {Kabuss}}, \bibinfo {author} {\bibfnamefont {A.}~\bibnamefont {Carmele}},
  \bibinfo {author} {\bibfnamefont {M.}~\bibnamefont {Richter}}, \ and\
  \bibinfo {author} {\bibfnamefont {A.}~\bibnamefont {Knorr}},\ }\href
  {\doibase 10.1103/PhysRevB.84.125324} {\bibfield  {journal} {\bibinfo
  {journal} {Phys. Rev. B}\ }\textbf {\bibinfo {volume} {84}},\ \bibinfo
  {pages} {125324} (\bibinfo {year} {2011}{\natexlab{a}})}\BibitemShut
  {NoStop}%
\bibitem [{\citenamefont {Kabuss}\ \emph
  {et~al.}(2011{\natexlab{b}})\citenamefont {Kabuss}, \citenamefont {Carmele},
  \citenamefont {Richter}, \citenamefont {Chow},\ and\ \citenamefont
  {Knorr}}]{kabuss2011inductive}%
  \BibitemOpen
  \bibfield  {author} {\bibinfo {author} {\bibfnamefont {J.}~\bibnamefont
  {Kabuss}}, \bibinfo {author} {\bibfnamefont {A.}~\bibnamefont {Carmele}},
  \bibinfo {author} {\bibfnamefont {M.}~\bibnamefont {Richter}}, \bibinfo
  {author} {\bibfnamefont {W.~W.}\ \bibnamefont {Chow}}, \ and\ \bibinfo
  {author} {\bibfnamefont {A.}~\bibnamefont {Knorr}},\ }\href@noop {}
  {\bibfield  {journal} {\bibinfo  {journal} {physica status solidi (b)}\
  }\textbf {\bibinfo {volume} {248}},\ \bibinfo {pages} {872} (\bibinfo {year}
  {2011}{\natexlab{b}})}\BibitemShut {NoStop}%
\bibitem [{\citenamefont {Naesby}\ \emph {et~al.}(2008)\citenamefont {Naesby},
  \citenamefont {Suhr}, \citenamefont {Kristensen},\ and\ \citenamefont
  {M\o{}rk}}]{Naesby2008}%
  \BibitemOpen
  \bibfield  {author} {\bibinfo {author} {\bibfnamefont {A.}~\bibnamefont
  {Naesby}}, \bibinfo {author} {\bibfnamefont {T.}~\bibnamefont {Suhr}},
  \bibinfo {author} {\bibfnamefont {P.~T.}\ \bibnamefont {Kristensen}}, \ and\
  \bibinfo {author} {\bibfnamefont {J.}~\bibnamefont {M\o{}rk}},\ }\href
  {\doibase 10.1103/PhysRevA.78.045802} {\bibfield  {journal} {\bibinfo
  {journal} {Phys. Rev. A}\ }\textbf {\bibinfo {volume} {78}},\ \bibinfo
  {pages} {045802} (\bibinfo {year} {2008})}\BibitemShut {NoStop}%
\bibitem [{\citenamefont {Reigue}\ \emph {et~al.}(2017)\citenamefont {Reigue},
  \citenamefont {Iles-Smith}, \citenamefont {Lux}, \citenamefont {Monniello},
  \citenamefont {Bernard}, \citenamefont {Margaillan}, \citenamefont
  {Lemaitre}, \citenamefont {Martinez}, \citenamefont {McCutcheon},
  \citenamefont {M\o{}rk}, \citenamefont {Hostein},\ and\ \citenamefont
  {Voliotis}}]{Reigue2017}%
  \BibitemOpen
  \bibfield  {author} {\bibinfo {author} {\bibfnamefont {A.}~\bibnamefont
  {Reigue}}, \bibinfo {author} {\bibfnamefont {J.}~\bibnamefont {Iles-Smith}},
  \bibinfo {author} {\bibfnamefont {F.}~\bibnamefont {Lux}}, \bibinfo {author}
  {\bibfnamefont {L.}~\bibnamefont {Monniello}}, \bibinfo {author}
  {\bibfnamefont {M.}~\bibnamefont {Bernard}}, \bibinfo {author} {\bibfnamefont
  {F.}~\bibnamefont {Margaillan}}, \bibinfo {author} {\bibfnamefont
  {A.}~\bibnamefont {Lemaitre}}, \bibinfo {author} {\bibfnamefont
  {A.}~\bibnamefont {Martinez}}, \bibinfo {author} {\bibfnamefont {D.~P.~S.}\
  \bibnamefont {McCutcheon}}, \bibinfo {author} {\bibfnamefont
  {J.}~\bibnamefont {M\o{}rk}}, \bibinfo {author} {\bibfnamefont
  {R.}~\bibnamefont {Hostein}}, \ and\ \bibinfo {author} {\bibfnamefont
  {V.}~\bibnamefont {Voliotis}},\ }\href {\doibase
  10.1103/PhysRevLett.118.233602} {\bibfield  {journal} {\bibinfo  {journal}
  {Phys. Rev. Lett.}\ }\textbf {\bibinfo {volume} {118}},\ \bibinfo {pages}
  {233602} (\bibinfo {year} {2017})}\BibitemShut {NoStop}%
\bibitem [{\citenamefont {Kreinberg}\ \emph {et~al.}(2018)\citenamefont
  {Kreinberg}, \citenamefont {Grbe{\v{s}}i{\'c}}, \citenamefont {Strau{\ss}},
  \citenamefont {Carmele}, \citenamefont {Emmerling}, \citenamefont
  {Schneider}, \citenamefont {H{\"o}fling}, \citenamefont {Porte},\ and\
  \citenamefont {Reitzenstein}}]{kreinberg2018quantum}%
  \BibitemOpen
  \bibfield  {author} {\bibinfo {author} {\bibfnamefont {S.}~\bibnamefont
  {Kreinberg}}, \bibinfo {author} {\bibfnamefont {T.}~\bibnamefont
  {Grbe{\v{s}}i{\'c}}}, \bibinfo {author} {\bibfnamefont {M.}~\bibnamefont
  {Strau{\ss}}}, \bibinfo {author} {\bibfnamefont {A.}~\bibnamefont {Carmele}},
  \bibinfo {author} {\bibfnamefont {M.}~\bibnamefont {Emmerling}}, \bibinfo
  {author} {\bibfnamefont {C.}~\bibnamefont {Schneider}}, \bibinfo {author}
  {\bibfnamefont {S.}~\bibnamefont {H{\"o}fling}}, \bibinfo {author}
  {\bibfnamefont {X.}~\bibnamefont {Porte}}, \ and\ \bibinfo {author}
  {\bibfnamefont {S.}~\bibnamefont {Reitzenstein}},\ }\href@noop {} {\bibfield
  {journal} {\bibinfo  {journal} {Light: Science \& Applications}\ }\textbf
  {\bibinfo {volume} {7}},\ \bibinfo {pages} {1} (\bibinfo {year}
  {2018})}\BibitemShut {NoStop}%
\bibitem [{\citenamefont {Ardelt}\ \emph {et~al.}(2014)\citenamefont {Ardelt},
  \citenamefont {Hanschke}, \citenamefont {Fischer}, \citenamefont {M\"uller},
  \citenamefont {Kleinkauf}, \citenamefont {Koller}, \citenamefont {Bechtold},
  \citenamefont {Simmet}, \citenamefont {Wierzbowski}, \citenamefont {Riedl},
  \citenamefont {Abstreiter},\ and\ \citenamefont {Finley}}]{Ardelt2014}%
  \BibitemOpen
  \bibfield  {author} {\bibinfo {author} {\bibfnamefont {P.-L.}\ \bibnamefont
  {Ardelt}}, \bibinfo {author} {\bibfnamefont {L.}~\bibnamefont {Hanschke}},
  \bibinfo {author} {\bibfnamefont {K.~A.}\ \bibnamefont {Fischer}}, \bibinfo
  {author} {\bibfnamefont {K.}~\bibnamefont {M\"uller}}, \bibinfo {author}
  {\bibfnamefont {A.}~\bibnamefont {Kleinkauf}}, \bibinfo {author}
  {\bibfnamefont {M.}~\bibnamefont {Koller}}, \bibinfo {author} {\bibfnamefont
  {A.}~\bibnamefont {Bechtold}}, \bibinfo {author} {\bibfnamefont
  {T.}~\bibnamefont {Simmet}}, \bibinfo {author} {\bibfnamefont
  {J.}~\bibnamefont {Wierzbowski}}, \bibinfo {author} {\bibfnamefont
  {H.}~\bibnamefont {Riedl}}, \bibinfo {author} {\bibfnamefont
  {G.}~\bibnamefont {Abstreiter}}, \ and\ \bibinfo {author} {\bibfnamefont
  {J.~J.}\ \bibnamefont {Finley}},\ }\href {\doibase
  10.1103/PhysRevB.90.241404} {\bibfield  {journal} {\bibinfo  {journal} {Phys.
  Rev. B}\ }\textbf {\bibinfo {volume} {90}},\ \bibinfo {pages} {241404}
  (\bibinfo {year} {2014})}\BibitemShut {NoStop}%
\bibitem [{\citenamefont {Gl\"assl}\ \emph {et~al.}(2011)\citenamefont
  {Gl\"assl}, \citenamefont {Vagov}, \citenamefont {L\"uker}, \citenamefont
  {Reiter}, \citenamefont {Croitoru}, \citenamefont {Machnikowski},
  \citenamefont {Axt},\ and\ \citenamefont {Kuhn}}]{Glaessl2011}%
  \BibitemOpen
  \bibfield  {author} {\bibinfo {author} {\bibfnamefont {M.}~\bibnamefont
  {Gl\"assl}}, \bibinfo {author} {\bibfnamefont {A.}~\bibnamefont {Vagov}},
  \bibinfo {author} {\bibfnamefont {S.}~\bibnamefont {L\"uker}}, \bibinfo
  {author} {\bibfnamefont {D.~E.}\ \bibnamefont {Reiter}}, \bibinfo {author}
  {\bibfnamefont {M.~D.}\ \bibnamefont {Croitoru}}, \bibinfo {author}
  {\bibfnamefont {P.}~\bibnamefont {Machnikowski}}, \bibinfo {author}
  {\bibfnamefont {V.~M.}\ \bibnamefont {Axt}}, \ and\ \bibinfo {author}
  {\bibfnamefont {T.}~\bibnamefont {Kuhn}},\ }\href {\doibase
  10.1103/PhysRevB.84.195311} {\bibfield  {journal} {\bibinfo  {journal} {Phys.
  Rev. B}\ }\textbf {\bibinfo {volume} {84}},\ \bibinfo {pages} {195311}
  (\bibinfo {year} {2011})}\BibitemShut {NoStop}%
\bibitem [{\citenamefont {Gl\"assl}\ \emph {et~al.}(2013)\citenamefont
  {Gl\"assl}, \citenamefont {Barth},\ and\ \citenamefont {Axt}}]{Glaessl2013}%
  \BibitemOpen
  \bibfield  {author} {\bibinfo {author} {\bibfnamefont {M.}~\bibnamefont
  {Gl\"assl}}, \bibinfo {author} {\bibfnamefont {A.~M.}\ \bibnamefont {Barth}},
  \ and\ \bibinfo {author} {\bibfnamefont {V.~M.}\ \bibnamefont {Axt}},\ }\href
  {\doibase 10.1103/PhysRevLett.110.147401} {\bibfield  {journal} {\bibinfo
  {journal} {Phys. Rev. Lett.}\ }\textbf {\bibinfo {volume} {110}},\ \bibinfo
  {pages} {147401} (\bibinfo {year} {2013})}\BibitemShut {NoStop}%
\bibitem [{\citenamefont {Barth}\ \emph
  {et~al.}(2016{\natexlab{a}})\citenamefont {Barth}, \citenamefont {L\"uker},
  \citenamefont {Vagov}, \citenamefont {Reiter}, \citenamefont {Kuhn},\ and\
  \citenamefont {Axt}}]{Barth2016fast}%
  \BibitemOpen
  \bibfield  {author} {\bibinfo {author} {\bibfnamefont {A.~M.}\ \bibnamefont
  {Barth}}, \bibinfo {author} {\bibfnamefont {S.}~\bibnamefont {L\"uker}},
  \bibinfo {author} {\bibfnamefont {A.}~\bibnamefont {Vagov}}, \bibinfo
  {author} {\bibfnamefont {D.~E.}\ \bibnamefont {Reiter}}, \bibinfo {author}
  {\bibfnamefont {T.}~\bibnamefont {Kuhn}}, \ and\ \bibinfo {author}
  {\bibfnamefont {V.~M.}\ \bibnamefont {Axt}},\ }\href {\doibase
  10.1103/PhysRevB.94.045306} {\bibfield  {journal} {\bibinfo  {journal} {Phys.
  Rev. B}\ }\textbf {\bibinfo {volume} {94}},\ \bibinfo {pages} {045306}
  (\bibinfo {year} {2016}{\natexlab{a}})}\BibitemShut {NoStop}%
\bibitem [{\citenamefont {Barth}\ \emph
  {et~al.}(2016{\natexlab{b}})\citenamefont {Barth}, \citenamefont {Vagov},\
  and\ \citenamefont {Axt}}]{Barth2016path}%
  \BibitemOpen
  \bibfield  {author} {\bibinfo {author} {\bibfnamefont {A.~M.}\ \bibnamefont
  {Barth}}, \bibinfo {author} {\bibfnamefont {A.}~\bibnamefont {Vagov}}, \ and\
  \bibinfo {author} {\bibfnamefont {V.~M.}\ \bibnamefont {Axt}},\ }\href
  {\doibase 10.1103/PhysRevB.94.125439} {\bibfield  {journal} {\bibinfo
  {journal} {Phys. Rev. B}\ }\textbf {\bibinfo {volume} {94}},\ \bibinfo
  {pages} {125439} (\bibinfo {year} {2016}{\natexlab{b}})}\BibitemShut
  {NoStop}%
\bibitem [{\citenamefont {Wu}\ \emph {et~al.}(2011)\citenamefont {Wu},
  \citenamefont {Piper}, \citenamefont {Ediger}, \citenamefont {Brereton},
  \citenamefont {Schmidgall}, \citenamefont {Eastham}, \citenamefont {Hugues},
  \citenamefont {Hopkinson},\ and\ \citenamefont {Phillips}}]{Wu2011}%
  \BibitemOpen
  \bibfield  {author} {\bibinfo {author} {\bibfnamefont {Y.}~\bibnamefont
  {Wu}}, \bibinfo {author} {\bibfnamefont {I.~M.}\ \bibnamefont {Piper}},
  \bibinfo {author} {\bibfnamefont {M.}~\bibnamefont {Ediger}}, \bibinfo
  {author} {\bibfnamefont {P.}~\bibnamefont {Brereton}}, \bibinfo {author}
  {\bibfnamefont {E.~R.}\ \bibnamefont {Schmidgall}}, \bibinfo {author}
  {\bibfnamefont {P.~R.}\ \bibnamefont {Eastham}}, \bibinfo {author}
  {\bibfnamefont {M.}~\bibnamefont {Hugues}}, \bibinfo {author} {\bibfnamefont
  {M.}~\bibnamefont {Hopkinson}}, \ and\ \bibinfo {author} {\bibfnamefont
  {R.~T.}\ \bibnamefont {Phillips}},\ }\href {\doibase
  10.1103/PhysRevLett.106.067401} {\bibfield  {journal} {\bibinfo  {journal}
  {Phys. Rev. Lett.}\ }\textbf {\bibinfo {volume} {106}},\ \bibinfo {pages}
  {067401} (\bibinfo {year} {2011})}\BibitemShut {NoStop}%
\bibitem [{\citenamefont {Paspalakis}\ \emph {et~al.}(2013)\citenamefont
  {Paspalakis}, \citenamefont {Evangelou},\ and\ \citenamefont
  {Terzis}}]{Paspalakis2013}%
  \BibitemOpen
  \bibfield  {author} {\bibinfo {author} {\bibfnamefont {E.}~\bibnamefont
  {Paspalakis}}, \bibinfo {author} {\bibfnamefont {S.}~\bibnamefont
  {Evangelou}}, \ and\ \bibinfo {author} {\bibfnamefont {A.~F.}\ \bibnamefont
  {Terzis}},\ }\href {\doibase 10.1103/PhysRevB.87.235302} {\bibfield
  {journal} {\bibinfo  {journal} {Phys. Rev. B}\ }\textbf {\bibinfo {volume}
  {87}},\ \bibinfo {pages} {235302} (\bibinfo {year} {2013})}\BibitemShut
  {NoStop}%
\bibitem [{\citenamefont {Roy}\ and\ \citenamefont {Hughes}(2011)}]{Roy2011}%
  \BibitemOpen
  \bibfield  {author} {\bibinfo {author} {\bibfnamefont {C.}~\bibnamefont
  {Roy}}\ and\ \bibinfo {author} {\bibfnamefont {S.}~\bibnamefont {Hughes}},\
  }\href {\doibase 10.1103/PhysRevX.1.021009} {\bibfield  {journal} {\bibinfo
  {journal} {Phys. Rev. X}\ }\textbf {\bibinfo {volume} {1}},\ \bibinfo {pages}
  {021009} (\bibinfo {year} {2011})}\BibitemShut {NoStop}%
\bibitem [{\citenamefont {Hughes}\ and\ \citenamefont
  {Carmichael}(2013)}]{Hughes2013}%
  \BibitemOpen
  \bibfield  {author} {\bibinfo {author} {\bibfnamefont {S.}~\bibnamefont
  {Hughes}}\ and\ \bibinfo {author} {\bibfnamefont {H.~J.}\ \bibnamefont
  {Carmichael}},\ }\href {\doibase 10.1088/1367-2630/15/5/053039} {\bibfield
  {journal} {\bibinfo  {journal} {New Journal of Physics}\ }\textbf {\bibinfo
  {volume} {15}},\ \bibinfo {pages} {053039} (\bibinfo {year}
  {2013})}\BibitemShut {NoStop}%
\bibitem [{\citenamefont {Metelmann}\ and\ \citenamefont
  {Clerk}(2015)}]{Metelmann2015}%
  \BibitemOpen
  \bibfield  {author} {\bibinfo {author} {\bibfnamefont {A.}~\bibnamefont
  {Metelmann}}\ and\ \bibinfo {author} {\bibfnamefont {A.~A.}\ \bibnamefont
  {Clerk}},\ }\href {\doibase 10.1103/PhysRevX.5.021025} {\bibfield  {journal}
  {\bibinfo  {journal} {Phys. Rev. X}\ }\textbf {\bibinfo {volume} {5}},\
  \bibinfo {pages} {021025} (\bibinfo {year} {2015})}\BibitemShut {NoStop}%
\bibitem [{\citenamefont {Quilter}\ \emph {et~al.}(2015)\citenamefont
  {Quilter}, \citenamefont {Brash}, \citenamefont {Liu}, \citenamefont
  {Gl\"assl}, \citenamefont {Barth}, \citenamefont {Axt}, \citenamefont
  {Ramsay}, \citenamefont {Skolnick},\ and\ \citenamefont {Fox}}]{Quilter2015}%
  \BibitemOpen
  \bibfield  {author} {\bibinfo {author} {\bibfnamefont {J.~H.}\ \bibnamefont
  {Quilter}}, \bibinfo {author} {\bibfnamefont {A.~J.}\ \bibnamefont {Brash}},
  \bibinfo {author} {\bibfnamefont {F.}~\bibnamefont {Liu}}, \bibinfo {author}
  {\bibfnamefont {M.}~\bibnamefont {Gl\"assl}}, \bibinfo {author}
  {\bibfnamefont {A.~M.}\ \bibnamefont {Barth}}, \bibinfo {author}
  {\bibfnamefont {V.~M.}\ \bibnamefont {Axt}}, \bibinfo {author} {\bibfnamefont
  {A.~J.}\ \bibnamefont {Ramsay}}, \bibinfo {author} {\bibfnamefont {M.~S.}\
  \bibnamefont {Skolnick}}, \ and\ \bibinfo {author} {\bibfnamefont {A.~M.}\
  \bibnamefont {Fox}},\ }\href {\doibase 10.1103/PhysRevLett.114.137401}
  {\bibfield  {journal} {\bibinfo  {journal} {Phys. Rev. Lett.}\ }\textbf
  {\bibinfo {volume} {114}},\ \bibinfo {pages} {137401} (\bibinfo {year}
  {2015})}\BibitemShut {NoStop}%
\bibitem [{\citenamefont {Thomas}\ \emph {et~al.}(2020)\citenamefont {Thomas},
  \citenamefont {Billard}, \citenamefont {Coste}, \citenamefont {Wein},
  \citenamefont {Priya}, \citenamefont {Ollivier}, \citenamefont {Krebs},
  \citenamefont {Tazaïrt}, \citenamefont {Harouri}, \citenamefont {Lemaitre},
  \citenamefont {Sagnes}, \citenamefont {Anton}, \citenamefont {Lanco},
  \citenamefont {Somaschi}, \citenamefont {Loredo},\ and\ \citenamefont
  {Senellart}}]{Thomas2020}%
  \BibitemOpen
  \bibfield  {author} {\bibinfo {author} {\bibfnamefont {S.~E.}\ \bibnamefont
  {Thomas}}, \bibinfo {author} {\bibfnamefont {M.}~\bibnamefont {Billard}},
  \bibinfo {author} {\bibfnamefont {N.}~\bibnamefont {Coste}}, \bibinfo
  {author} {\bibfnamefont {S.~C.}\ \bibnamefont {Wein}}, \bibinfo {author}
  {\bibnamefont {Priya}}, \bibinfo {author} {\bibfnamefont {H.}~\bibnamefont
  {Ollivier}}, \bibinfo {author} {\bibfnamefont {O.}~\bibnamefont {Krebs}},
  \bibinfo {author} {\bibfnamefont {L.}~\bibnamefont {Tazaïrt}}, \bibinfo
  {author} {\bibfnamefont {A.}~\bibnamefont {Harouri}}, \bibinfo {author}
  {\bibfnamefont {A.}~\bibnamefont {Lemaitre}}, \bibinfo {author}
  {\bibfnamefont {I.}~\bibnamefont {Sagnes}}, \bibinfo {author} {\bibfnamefont
  {C.}~\bibnamefont {Anton}}, \bibinfo {author} {\bibfnamefont
  {L.}~\bibnamefont {Lanco}}, \bibinfo {author} {\bibfnamefont
  {N.}~\bibnamefont {Somaschi}}, \bibinfo {author} {\bibfnamefont {J.~C.}\
  \bibnamefont {Loredo}}, \ and\ \bibinfo {author} {\bibfnamefont
  {P.}~\bibnamefont {Senellart}},\ }\href@noop {} {\bibfield  {journal}
  {\bibinfo  {journal} {arXiv:2007.04330}\ } (\bibinfo {year}
  {2020})}\BibitemShut {NoStop}%
\bibitem [{\citenamefont {Houmark}\ \emph {et~al.}(2009)\citenamefont
  {Houmark}, \citenamefont {Nielsen}, \citenamefont {M\o{}rk},\ and\
  \citenamefont {Jauho}}]{Houmark2009}%
  \BibitemOpen
  \bibfield  {author} {\bibinfo {author} {\bibfnamefont {J.}~\bibnamefont
  {Houmark}}, \bibinfo {author} {\bibfnamefont {T.~R.}\ \bibnamefont
  {Nielsen}}, \bibinfo {author} {\bibfnamefont {J.}~\bibnamefont {M\o{}rk}}, \
  and\ \bibinfo {author} {\bibfnamefont {A.-P.}\ \bibnamefont {Jauho}},\ }\href
  {\doibase 10.1103/PhysRevB.79.115420} {\bibfield  {journal} {\bibinfo
  {journal} {Phys. Rev. B}\ }\textbf {\bibinfo {volume} {79}},\ \bibinfo
  {pages} {115420} (\bibinfo {year} {2009})}\BibitemShut {NoStop}%
\bibitem [{\citenamefont {Barettin}\ \emph {et~al.}(2009)\citenamefont
  {Barettin}, \citenamefont {Houmark}, \citenamefont {Lassen}, \citenamefont
  {Willatzen}, \citenamefont {Nielsen}, \citenamefont {M\o{}rk},\ and\
  \citenamefont {Jauho}}]{Barettin2009}%
  \BibitemOpen
  \bibfield  {author} {\bibinfo {author} {\bibfnamefont {D.}~\bibnamefont
  {Barettin}}, \bibinfo {author} {\bibfnamefont {J.}~\bibnamefont {Houmark}},
  \bibinfo {author} {\bibfnamefont {B.}~\bibnamefont {Lassen}}, \bibinfo
  {author} {\bibfnamefont {M.}~\bibnamefont {Willatzen}}, \bibinfo {author}
  {\bibfnamefont {T.~R.}\ \bibnamefont {Nielsen}}, \bibinfo {author}
  {\bibfnamefont {J.}~\bibnamefont {M\o{}rk}}, \ and\ \bibinfo {author}
  {\bibfnamefont {A.-P.}\ \bibnamefont {Jauho}},\ }\href {\doibase
  10.1103/PhysRevB.80.235304} {\bibfield  {journal} {\bibinfo  {journal} {Phys.
  Rev. B}\ }\textbf {\bibinfo {volume} {80}},\ \bibinfo {pages} {235304}
  (\bibinfo {year} {2009})}\BibitemShut {NoStop}%
\bibitem [{\citenamefont {Wang}\ \emph {et~al.}(2005)\citenamefont {Wang},
  \citenamefont {Muller}, \citenamefont {Cheng}, \citenamefont {Zhou},
  \citenamefont {Bianucci},\ and\ \citenamefont {Shih}}]{Wang2005}%
  \BibitemOpen
  \bibfield  {author} {\bibinfo {author} {\bibfnamefont {Q.~Q.}\ \bibnamefont
  {Wang}}, \bibinfo {author} {\bibfnamefont {A.}~\bibnamefont {Muller}},
  \bibinfo {author} {\bibfnamefont {M.~T.}\ \bibnamefont {Cheng}}, \bibinfo
  {author} {\bibfnamefont {H.~J.}\ \bibnamefont {Zhou}}, \bibinfo {author}
  {\bibfnamefont {P.}~\bibnamefont {Bianucci}}, \ and\ \bibinfo {author}
  {\bibfnamefont {C.~K.}\ \bibnamefont {Shih}},\ }\href {\doibase
  10.1103/PhysRevLett.95.187404} {\bibfield  {journal} {\bibinfo  {journal}
  {Phys. Rev. Lett.}\ }\textbf {\bibinfo {volume} {95}},\ \bibinfo {pages}
  {187404} (\bibinfo {year} {2005})}\BibitemShut {NoStop}%
\bibitem [{\citenamefont {Breuer}\ and\ \citenamefont
  {Petruccione}(2002)}]{Breuer2002}%
  \BibitemOpen
  \bibfield  {author} {\bibinfo {author} {\bibfnamefont {H.~P.}\ \bibnamefont
  {Breuer}}\ and\ \bibinfo {author} {\bibfnamefont {F.}~\bibnamefont
  {Petruccione}},\ }\href@noop {} {\emph {\bibinfo {title} {The theory of open
  quantum systems}}}\ (\bibinfo  {publisher} {Oxford University Press},\
  \bibinfo {year} {2002})\BibitemShut {NoStop}%
\bibitem [{\citenamefont {de~Vega}\ and\ \citenamefont
  {Alonso}(2017)}]{Vega2017}%
  \BibitemOpen
  \bibfield  {author} {\bibinfo {author} {\bibfnamefont {I.}~\bibnamefont
  {de~Vega}}\ and\ \bibinfo {author} {\bibfnamefont {D.}~\bibnamefont
  {Alonso}},\ }\href {\doibase 10.1103/RevModPhys.89.015001} {\bibfield
  {journal} {\bibinfo  {journal} {Rev. Mod. Phys.}\ }\textbf {\bibinfo {volume}
  {89}},\ \bibinfo {pages} {015001} (\bibinfo {year} {2017})}\BibitemShut
  {NoStop}%
\bibitem [{\citenamefont {Carmele}\ and\ \citenamefont
  {Reitzenstein}(2019)}]{Carmele2019}%
  \BibitemOpen
  \bibfield  {author} {\bibinfo {author} {\bibfnamefont {A.}~\bibnamefont
  {Carmele}}\ and\ \bibinfo {author} {\bibfnamefont {S.}~\bibnamefont
  {Reitzenstein}},\ }\href {\doibase 10.1515/nanoph-2018-0222} {\bibfield
  {journal} {\bibinfo  {journal} {Nanophotonics}\ }\textbf {\bibinfo {volume}
  {8}},\ \bibinfo {pages} {655} (\bibinfo {year} {2019})}\BibitemShut {NoStop}%
\bibitem [{\citenamefont {Beenakker}\ and\ \citenamefont {van
  Houten}(1991)}]{Beenakker1991}%
  \BibitemOpen
  \bibfield  {author} {\bibinfo {author} {\bibfnamefont {C.}~\bibnamefont
  {Beenakker}}\ and\ \bibinfo {author} {\bibfnamefont {H.}~\bibnamefont {van
  Houten}},\ }in\ \href {\doibase
  https://doi.org/10.1016/S0081-1947(08)60091-0} {\emph {\bibinfo {booktitle}
  {Semiconductor Heterostructures and Nanostructures}}},\ \bibinfo {series}
  {Solid State Physics}, Vol.~\bibinfo {volume} {44},\ \bibinfo {editor}
  {edited by\ \bibinfo {editor} {\bibfnamefont {H.}~\bibnamefont {Ehrenreich}}\
  and\ \bibinfo {editor} {\bibfnamefont {D.}~\bibnamefont {Turnbull}}}\
  (\bibinfo  {publisher} {Academic Press},\ \bibinfo {year} {1991})\BibitemShut
  {NoStop}%
\bibitem [{\citenamefont {Crooker}\ \emph {et~al.}(2002)\citenamefont
  {Crooker}, \citenamefont {Hollingsworth}, \citenamefont {Tretiak},\ and\
  \citenamefont {Klimov}}]{Crooker2002}%
  \BibitemOpen
  \bibfield  {author} {\bibinfo {author} {\bibfnamefont {S.~A.}\ \bibnamefont
  {Crooker}}, \bibinfo {author} {\bibfnamefont {J.~A.}\ \bibnamefont
  {Hollingsworth}}, \bibinfo {author} {\bibfnamefont {S.}~\bibnamefont
  {Tretiak}}, \ and\ \bibinfo {author} {\bibfnamefont {V.~I.}\ \bibnamefont
  {Klimov}},\ }\href {\doibase 10.1103/PhysRevLett.89.186802} {\bibfield
  {journal} {\bibinfo  {journal} {Phys. Rev. Lett.}\ }\textbf {\bibinfo
  {volume} {89}},\ \bibinfo {pages} {186802} (\bibinfo {year}
  {2002})}\BibitemShut {NoStop}%
\bibitem [{\citenamefont {Machida}\ \emph {et~al.}(2003)\citenamefont
  {Machida}, \citenamefont {Matsuo}, \citenamefont {Fujiwara}, \citenamefont
  {Folkenberg},\ and\ \citenamefont {Hvam}}]{Machida2003}%
  \BibitemOpen
  \bibfield  {author} {\bibinfo {author} {\bibfnamefont {S.}~\bibnamefont
  {Machida}}, \bibinfo {author} {\bibfnamefont {M.}~\bibnamefont {Matsuo}},
  \bibinfo {author} {\bibfnamefont {K.}~\bibnamefont {Fujiwara}}, \bibinfo
  {author} {\bibfnamefont {J.~R.}\ \bibnamefont {Folkenberg}}, \ and\ \bibinfo
  {author} {\bibfnamefont {J.~M.}\ \bibnamefont {Hvam}},\ }\href {\doibase
  10.1103/PhysRevB.67.205322} {\bibfield  {journal} {\bibinfo  {journal} {Phys.
  Rev. B}\ }\textbf {\bibinfo {volume} {67}},\ \bibinfo {pages} {205322}
  (\bibinfo {year} {2003})}\BibitemShut {NoStop}%
\bibitem [{\citenamefont {Sangu}\ \emph {et~al.}(2003)\citenamefont {Sangu},
  \citenamefont {Kobayashi}, \citenamefont {Shojiguchi}, \citenamefont
  {Kawazoe},\ and\ \citenamefont {Ohtsu}}]{Sangu2003}%
  \BibitemOpen
  \bibfield  {author} {\bibinfo {author} {\bibfnamefont {S.}~\bibnamefont
  {Sangu}}, \bibinfo {author} {\bibfnamefont {K.}~\bibnamefont {Kobayashi}},
  \bibinfo {author} {\bibfnamefont {A.}~\bibnamefont {Shojiguchi}}, \bibinfo
  {author} {\bibfnamefont {T.}~\bibnamefont {Kawazoe}}, \ and\ \bibinfo
  {author} {\bibfnamefont {M.}~\bibnamefont {Ohtsu}},\ }\href {\doibase
  10.1063/1.1540739} {\bibfield  {journal} {\bibinfo  {journal} {Journal of
  Applied Physics}\ }\textbf {\bibinfo {volume} {93}},\ \bibinfo {pages} {2937}
  (\bibinfo {year} {2003})}\BibitemShut {NoStop}%
\bibitem [{\citenamefont {Lu}\ and\ \citenamefont {Madhukar}(2007)}]{Lu2007}%
  \BibitemOpen
  \bibfield  {author} {\bibinfo {author} {\bibfnamefont {S.}~\bibnamefont
  {Lu}}\ and\ \bibinfo {author} {\bibfnamefont {A.}~\bibnamefont {Madhukar}},\
  }\href {\doibase 10.1021/nl0719731} {\bibfield  {journal} {\bibinfo
  {journal} {Nano Letters}\ }\textbf {\bibinfo {volume} {7}},\ \bibinfo {pages}
  {3443} (\bibinfo {year} {2007})}\BibitemShut {NoStop}%
\bibitem [{\citenamefont {Chandler}\ \emph {et~al.}(2007)\citenamefont
  {Chandler}, \citenamefont {Houtepen}, \citenamefont {Nelson},\ and\
  \citenamefont {Vanmaekelbergh}}]{Chandler2007}%
  \BibitemOpen
  \bibfield  {author} {\bibinfo {author} {\bibfnamefont {R.~E.}\ \bibnamefont
  {Chandler}}, \bibinfo {author} {\bibfnamefont {A.~J.}\ \bibnamefont
  {Houtepen}}, \bibinfo {author} {\bibfnamefont {J.}~\bibnamefont {Nelson}}, \
  and\ \bibinfo {author} {\bibfnamefont {D.}~\bibnamefont {Vanmaekelbergh}},\
  }\href {\doibase 10.1103/PhysRevB.75.085325} {\bibfield  {journal} {\bibinfo
  {journal} {Phys. Rev. B}\ }\textbf {\bibinfo {volume} {75}},\ \bibinfo
  {pages} {085325} (\bibinfo {year} {2007})}\BibitemShut {NoStop}%
\bibitem [{\citenamefont {Iotti}\ \emph {et~al.}(2005)\citenamefont {Iotti},
  \citenamefont {Ciancio},\ and\ \citenamefont {Rossi}}]{Iotti2005}%
  \BibitemOpen
  \bibfield  {author} {\bibinfo {author} {\bibfnamefont {R.~C.}\ \bibnamefont
  {Iotti}}, \bibinfo {author} {\bibfnamefont {E.}~\bibnamefont {Ciancio}}, \
  and\ \bibinfo {author} {\bibfnamefont {F.}~\bibnamefont {Rossi}},\ }\href
  {\doibase 10.1103/PhysRevB.72.125347} {\bibfield  {journal} {\bibinfo
  {journal} {Phys. Rev. B}\ }\textbf {\bibinfo {volume} {72}},\ \bibinfo
  {pages} {125347} (\bibinfo {year} {2005})}\BibitemShut {NoStop}%
\bibitem [{\citenamefont {Richter}\ \emph {et~al.}(2006)\citenamefont
  {Richter}, \citenamefont {Ahn}, \citenamefont {Knorr}, \citenamefont
  {Schliwa}, \citenamefont {Bimberg}, \citenamefont {Madjet},\ and\
  \citenamefont {Renger}}]{Richter2006}%
  \BibitemOpen
  \bibfield  {author} {\bibinfo {author} {\bibfnamefont {M.}~\bibnamefont
  {Richter}}, \bibinfo {author} {\bibfnamefont {K.~J.}\ \bibnamefont {Ahn}},
  \bibinfo {author} {\bibfnamefont {A.}~\bibnamefont {Knorr}}, \bibinfo
  {author} {\bibfnamefont {A.}~\bibnamefont {Schliwa}}, \bibinfo {author}
  {\bibfnamefont {D.}~\bibnamefont {Bimberg}}, \bibinfo {author} {\bibfnamefont
  {M.~E.-A.}\ \bibnamefont {Madjet}}, \ and\ \bibinfo {author} {\bibfnamefont
  {T.}~\bibnamefont {Renger}},\ }\href {\doibase 10.1002/pssb.200668053}
  {\bibfield  {journal} {\bibinfo  {journal} {physica status solidi (b)}\
  }\textbf {\bibinfo {volume} {243}},\ \bibinfo {pages} {2302} (\bibinfo {year}
  {2006})}\BibitemShut {NoStop}%
\bibitem [{\citenamefont {M\'etivier}\ \emph {et~al.}(2007)\citenamefont
  {M\'etivier}, \citenamefont {Nolde}, \citenamefont {M\"ullen},\ and\
  \citenamefont {Basch\'e}}]{Metivier2007}%
  \BibitemOpen
  \bibfield  {author} {\bibinfo {author} {\bibfnamefont {R.}~\bibnamefont
  {M\'etivier}}, \bibinfo {author} {\bibfnamefont {F.}~\bibnamefont {Nolde}},
  \bibinfo {author} {\bibfnamefont {K.}~\bibnamefont {M\"ullen}}, \ and\
  \bibinfo {author} {\bibfnamefont {T.}~\bibnamefont {Basch\'e}},\ }\href
  {\doibase 10.1103/PhysRevLett.98.047802} {\bibfield  {journal} {\bibinfo
  {journal} {Phys. Rev. Lett.}\ }\textbf {\bibinfo {volume} {98}},\ \bibinfo
  {pages} {047802} (\bibinfo {year} {2007})}\BibitemShut {NoStop}%
\bibitem [{\citenamefont {Leitner}\ and\ \citenamefont
  {Pandey}(2015)}]{Leitner2015}%
  \BibitemOpen
  \bibfield  {author} {\bibinfo {author} {\bibfnamefont {D.~M.}\ \bibnamefont
  {Leitner}}\ and\ \bibinfo {author} {\bibfnamefont {H.~D.}\ \bibnamefont
  {Pandey}},\ }\href {\doibase 10.1002/andp.201500104} {\bibfield  {journal}
  {\bibinfo  {journal} {Annalen der Physik}\ }\textbf {\bibinfo {volume}
  {527}},\ \bibinfo {pages} {601} (\bibinfo {year} {2015})}\BibitemShut
  {NoStop}%
\bibitem [{\citenamefont {Cupellini}\ \emph {et~al.}(2018)\citenamefont
  {Cupellini}, \citenamefont {Caprasecca}, \citenamefont {Guido}, \citenamefont
  {Müh}, \citenamefont {Renger},\ and\ \citenamefont
  {Mennucci}}]{Cupellini2018}%
  \BibitemOpen
  \bibfield  {author} {\bibinfo {author} {\bibfnamefont {L.}~\bibnamefont
  {Cupellini}}, \bibinfo {author} {\bibfnamefont {S.}~\bibnamefont
  {Caprasecca}}, \bibinfo {author} {\bibfnamefont {C.~A.}\ \bibnamefont
  {Guido}}, \bibinfo {author} {\bibfnamefont {F.}~\bibnamefont {Müh}},
  \bibinfo {author} {\bibfnamefont {T.}~\bibnamefont {Renger}}, \ and\ \bibinfo
  {author} {\bibfnamefont {B.}~\bibnamefont {Mennucci}},\ } {\bibfield  {journal} {\bibinfo  {journal} {The
  Journal of Physical Chemistry Letters}\ }\textbf {\bibinfo {volume} {9}},\
  \bibinfo {pages} {6892} (\bibinfo {year} {2018})}\BibitemShut {NoStop}%
\bibitem [{\citenamefont {Shangguan}\ \emph {et~al.}(2001)\citenamefont
  {Shangguan}, \citenamefont {Au~Yeung}, \citenamefont {Yu},\ and\
  \citenamefont {Kam}}]{Shangguan2001}%
  \BibitemOpen
  \bibfield  {author} {\bibinfo {author} {\bibfnamefont {W.~Z.}\ \bibnamefont
  {Shangguan}}, \bibinfo {author} {\bibfnamefont {T.~C.}\ \bibnamefont
  {Au~Yeung}}, \bibinfo {author} {\bibfnamefont {Y.~B.}\ \bibnamefont {Yu}}, \
  and\ \bibinfo {author} {\bibfnamefont {C.~H.}\ \bibnamefont {Kam}},\ }\href
  {\doibase 10.1103/PhysRevB.63.235323} {\bibfield  {journal} {\bibinfo
  {journal} {Phys. Rev. B}\ }\textbf {\bibinfo {volume} {63}},\ \bibinfo
  {pages} {235323} (\bibinfo {year} {2001})}\BibitemShut {NoStop}%
\bibitem [{\citenamefont {Zheng}\ \emph {et~al.}(2009)\citenamefont {Zheng},
  \citenamefont {Luo}, \citenamefont {Jin},\ and\ \citenamefont
  {Yan}}]{Zheng2009}%
  \BibitemOpen
  \bibfield  {author} {\bibinfo {author} {\bibfnamefont {X.}~\bibnamefont
  {Zheng}}, \bibinfo {author} {\bibfnamefont {J.}~\bibnamefont {Luo}}, \bibinfo
  {author} {\bibfnamefont {J.}~\bibnamefont {Jin}}, \ and\ \bibinfo {author}
  {\bibfnamefont {Y.}~\bibnamefont {Yan}},\ }\href {\doibase 10.1063/1.3095424}
  {\bibfield  {journal} {\bibinfo  {journal} {The Journal of Chemical Physics}\
  }\textbf {\bibinfo {volume} {130}},\ \bibinfo {pages} {124508} (\bibinfo
  {year} {2009})}\BibitemShut {NoStop}%
\bibitem [{\citenamefont {Rebentrost}\ \emph {et~al.}(2009)\citenamefont
  {Rebentrost}, \citenamefont {Mohseni}, \citenamefont {Kassal}, \citenamefont
  {Lloyd},\ and\ \citenamefont {Aspuru-Guzik}}]{Rebentrost2009}%
  \BibitemOpen
  \bibfield  {author} {\bibinfo {author} {\bibfnamefont {P.}~\bibnamefont
  {Rebentrost}}, \bibinfo {author} {\bibfnamefont {M.}~\bibnamefont {Mohseni}},
  \bibinfo {author} {\bibfnamefont {I.}~\bibnamefont {Kassal}}, \bibinfo
  {author} {\bibfnamefont {S.}~\bibnamefont {Lloyd}}, \ and\ \bibinfo {author}
  {\bibfnamefont {A.}~\bibnamefont {Aspuru-Guzik}},\ }\href {\doibase
  10.1088/1367-2630/11/3/033003} {\bibfield  {journal} {\bibinfo  {journal}
  {New Journal of Physics}\ }\textbf {\bibinfo {volume} {11}},\ \bibinfo
  {pages} {033003} (\bibinfo {year} {2009})}\BibitemShut {NoStop}%
\bibitem [{\citenamefont {W\"urger}(1998)}]{Wurger1998}%
  \BibitemOpen
  \bibfield  {author} {\bibinfo {author} {\bibfnamefont {A.}~\bibnamefont
  {W\"urger}},\ }\href {\doibase 10.1103/PhysRevB.57.347} {\bibfield  {journal}
  {\bibinfo  {journal} {Phys. Rev. B}\ }\textbf {\bibinfo {volume} {57}},\
  \bibinfo {pages} {347} (\bibinfo {year} {1998})}\BibitemShut {NoStop}%
\bibitem [{\citenamefont {Wilson-Rae}\ and\ \citenamefont
  {Imamo\ifmmode~\breve{g}\else \u{g}\fi{}lu}(2002)}]{Wilson2002}%
  \BibitemOpen
  \bibfield  {author} {\bibinfo {author} {\bibfnamefont {I.}~\bibnamefont
  {Wilson-Rae}}\ and\ \bibinfo {author} {\bibfnamefont {A.}~\bibnamefont
  {Imamo\ifmmode~\breve{g}\else \u{g}\fi{}lu}},\ }\href {\doibase
  10.1103/PhysRevB.65.235311} {\bibfield  {journal} {\bibinfo  {journal} {Phys.
  Rev. B}\ }\textbf {\bibinfo {volume} {65}},\ \bibinfo {pages} {235311}
  (\bibinfo {year} {2002})}\BibitemShut {NoStop}%
\bibitem [{\citenamefont {Manson}\ \emph {et~al.}(2016)\citenamefont {Manson},
  \citenamefont {Roy-Choudhury},\ and\ \citenamefont {Hughes}}]{Manson2016}%
  \BibitemOpen
  \bibfield  {author} {\bibinfo {author} {\bibfnamefont {R.}~\bibnamefont
  {Manson}}, \bibinfo {author} {\bibfnamefont {K.}~\bibnamefont
  {Roy-Choudhury}}, \ and\ \bibinfo {author} {\bibfnamefont {S.}~\bibnamefont
  {Hughes}},\ }\href {\doibase 10.1103/PhysRevB.93.155423} {\bibfield
  {journal} {\bibinfo  {journal} {Phys. Rev. B}\ }\textbf {\bibinfo {volume}
  {93}},\ \bibinfo {pages} {155423} (\bibinfo {year} {2016})}\BibitemShut
  {NoStop}%
\bibitem [{\citenamefont {Lee}\ \emph {et~al.}(2012)\citenamefont {Lee},
  \citenamefont {Moix},\ and\ \citenamefont {Cao}}]{Lee2012}%
  \BibitemOpen
  \bibfield  {author} {\bibinfo {author} {\bibfnamefont {C.~K.}\ \bibnamefont
  {Lee}}, \bibinfo {author} {\bibfnamefont {J.}~\bibnamefont {Moix}}, \ and\
  \bibinfo {author} {\bibfnamefont {J.}~\bibnamefont {Cao}},\ }\href {\doibase
  10.1063/1.4722336} {\bibfield  {journal} {\bibinfo  {journal} {The Journal of
  Chemical Physics}\ }\textbf {\bibinfo {volume} {136}},\ \bibinfo {pages}
  {204120} (\bibinfo {year} {2012})}\BibitemShut {NoStop}%
\bibitem [{\citenamefont {McCutcheon}\ and\ \citenamefont
  {Nazir}(2010)}]{McCutcheon2010}%
  \BibitemOpen
  \bibfield  {author} {\bibinfo {author} {\bibfnamefont {D.~P.~S.}\
  \bibnamefont {McCutcheon}}\ and\ \bibinfo {author} {\bibfnamefont
  {A.}~\bibnamefont {Nazir}},\ }\href {\doibase 10.1088/1367-2630/12/11/113042}
  {\bibfield  {journal} {\bibinfo  {journal} {New Journal of Physics}\ }\textbf
  {\bibinfo {volume} {12}},\ \bibinfo {pages} {113042} (\bibinfo {year}
  {2010})}\BibitemShut {NoStop}%
\bibitem [{\citenamefont {Jang}\ \emph {et~al.}(2008)\citenamefont {Jang},
  \citenamefont {Cheng}, \citenamefont {Reichman},\ and\ \citenamefont
  {Eaves}}]{Jang2008}%
  \BibitemOpen
  \bibfield  {author} {\bibinfo {author} {\bibfnamefont {S.}~\bibnamefont
  {Jang}}, \bibinfo {author} {\bibfnamefont {Y.-C.}\ \bibnamefont {Cheng}},
  \bibinfo {author} {\bibfnamefont {D.~R.}\ \bibnamefont {Reichman}}, \ and\
  \bibinfo {author} {\bibfnamefont {J.~D.}\ \bibnamefont {Eaves}},\ }\href@noop
  {} {\bibfield  {journal} {\bibinfo  {journal} {The Journal of Chemical
  Physics}\ }\textbf {\bibinfo {volume} {129}},\ \bibinfo {pages} {101104}
  (\bibinfo {year} {2008})}\BibitemShut {NoStop}%
\bibitem [{\citenamefont {Kolli}\ \emph {et~al.}(2011)\citenamefont {Kolli},
  \citenamefont {Nazir},\ and\ \citenamefont {Olaya-Castro}}]{Kolli2011}%
  \BibitemOpen
  \bibfield  {author} {\bibinfo {author} {\bibfnamefont {A.}~\bibnamefont
  {Kolli}}, \bibinfo {author} {\bibfnamefont {A.}~\bibnamefont {Nazir}}, \ and\
  \bibinfo {author} {\bibfnamefont {A.}~\bibnamefont {Olaya-Castro}},\
  }\href@noop {} {\bibfield  {journal} {\bibinfo  {journal} {The Journal of
  Chemical Physics}\ }\textbf {\bibinfo {volume} {135}},\ \bibinfo {pages}
  {154112} (\bibinfo {year} {2011})}\BibitemShut {NoStop}%
\bibitem [{\citenamefont {Chang}\ \emph {et~al.}(2013)\citenamefont {Chang},
  \citenamefont {Zhang},\ and\ \citenamefont {Cheng}}]{Chang2013}%
  \BibitemOpen
  \bibfield  {author} {\bibinfo {author} {\bibfnamefont {H.-T.}\ \bibnamefont
  {Chang}}, \bibinfo {author} {\bibfnamefont {P.-P.}\ \bibnamefont {Zhang}}, \
  and\ \bibinfo {author} {\bibfnamefont {Y.-C.}\ \bibnamefont {Cheng}},\
  }\href@noop {} {\bibfield  {journal} {\bibinfo  {journal} {The Journal of
  Chemical Physics}\ }\textbf {\bibinfo {volume} {139}},\ \bibinfo {pages}
  {224112} (\bibinfo {year} {2013})}\BibitemShut {NoStop}%
\bibitem [{\citenamefont {Dachner}\ \emph {et~al.}(2010)\citenamefont
  {Dachner}, \citenamefont {Malic}, \citenamefont {Richter}, \citenamefont
  {Carmele}, \citenamefont {Kabuss}, \citenamefont {Wilms}, \citenamefont
  {Kim}, \citenamefont {Hartmann}, \citenamefont {Wolters}, \citenamefont
  {Bandelow},\ and\ \citenamefont {Knorr}}]{Dachner2010}%
  \BibitemOpen
  \bibfield  {author} {\bibinfo {author} {\bibfnamefont {M.-R.}\ \bibnamefont
  {Dachner}}, \bibinfo {author} {\bibfnamefont {E.}~\bibnamefont {Malic}},
  \bibinfo {author} {\bibfnamefont {M.}~\bibnamefont {Richter}}, \bibinfo
  {author} {\bibfnamefont {A.}~\bibnamefont {Carmele}}, \bibinfo {author}
  {\bibfnamefont {J.}~\bibnamefont {Kabuss}}, \bibinfo {author} {\bibfnamefont
  {A.}~\bibnamefont {Wilms}}, \bibinfo {author} {\bibfnamefont {J.-E.}\
  \bibnamefont {Kim}}, \bibinfo {author} {\bibfnamefont {G.}~\bibnamefont
  {Hartmann}}, \bibinfo {author} {\bibfnamefont {J.}~\bibnamefont {Wolters}},
  \bibinfo {author} {\bibfnamefont {U.}~\bibnamefont {Bandelow}}, \ and\
  \bibinfo {author} {\bibfnamefont {A.}~\bibnamefont {Knorr}},\ }\href
  {\doibase 10.1002/pssb.200945433} {\bibfield  {journal} {\bibinfo  {journal}
  {physica status solidi (b)}\ }\textbf {\bibinfo {volume} {247}},\ \bibinfo
  {pages} {809} (\bibinfo {year} {2010})}\BibitemShut {NoStop}%
\bibitem [{\citenamefont {Melnik}\ and\ \citenamefont
  {Willatzen}(2003)}]{Melnik2003}%
  \BibitemOpen
  \bibfield  {author} {\bibinfo {author} {\bibfnamefont {R.~V.~N.}\
  \bibnamefont {Melnik}}\ and\ \bibinfo {author} {\bibfnamefont
  {M.}~\bibnamefont {Willatzen}},\ }\href {\doibase 10.1088/0957-4484/15/1/001}
  {\bibfield  {journal} {\bibinfo  {journal} {Nanotechnology}\ }\textbf
  {\bibinfo {volume} {15}},\ \bibinfo {pages} {1} (\bibinfo {year}
  {2003})}\BibitemShut {NoStop}%
\bibitem [{\citenamefont {Ahn}\ \emph {et~al.}(2005)\citenamefont {Ahn},
  \citenamefont {F\"orstner},\ and\ \citenamefont {Knorr}}]{Ahn2005}%
  \BibitemOpen
  \bibfield  {author} {\bibinfo {author} {\bibfnamefont {K.~J.}\ \bibnamefont
  {Ahn}}, \bibinfo {author} {\bibfnamefont {J.}~\bibnamefont {F\"orstner}}, \
  and\ \bibinfo {author} {\bibfnamefont {A.}~\bibnamefont {Knorr}},\ }\href
  {\doibase 10.1103/PhysRevB.71.153309} {\bibfield  {journal} {\bibinfo
  {journal} {Phys. Rev. B}\ }\textbf {\bibinfo {volume} {71}},\ \bibinfo
  {pages} {153309} (\bibinfo {year} {2005})}\BibitemShut {NoStop}%
\bibitem [{\citenamefont {Abrikosov}\ \emph {et~al.}(1965)\citenamefont
  {Abrikosov}, \citenamefont {Gorkov},\ and\ \citenamefont
  {Dzialoshinskii}}]{Abrikosov1965}%
  \BibitemOpen
  \bibfield  {author} {\bibinfo {author} {\bibfnamefont {A.~A.}\ \bibnamefont
  {Abrikosov}}, \bibinfo {author} {\bibfnamefont {L.~P.}\ \bibnamefont
  {Gorkov}}, \ and\ \bibinfo {author} {\bibfnamefont {I.~E.}\ \bibnamefont
  {Dzialoshinskii}},\ }\href@noop {} {\emph {\bibinfo {title} {Quantum Field
  Theoretical Methods in Statistical Physics}}}\ (\bibinfo  {publisher}
  {Pergamon},\ \bibinfo {year} {1965})\BibitemShut {NoStop}%
\bibitem [{\citenamefont {Mahan}(2000)}]{Mahan2000}%
  \BibitemOpen
  \bibfield  {author} {\bibinfo {author} {\bibfnamefont {G.~D.}\ \bibnamefont
  {Mahan}},\ }\href@noop {} {\emph {\bibinfo {title} {Many-Particle Physics}}}\
  (\bibinfo  {publisher} {Kluwer Academic/Plenum},\ \bibinfo {year}
  {2000})\BibitemShut {NoStop}%
\bibitem [{\citenamefont {F\"orstner}\ \emph {et~al.}(2003)\citenamefont
  {F\"orstner}, \citenamefont {Weber}, \citenamefont {Danckwerts},\ and\
  \citenamefont {Knorr}}]{Foerstner2003}%
  \BibitemOpen
  \bibfield  {author} {\bibinfo {author} {\bibfnamefont {J.}~\bibnamefont
  {F\"orstner}}, \bibinfo {author} {\bibfnamefont {C.}~\bibnamefont {Weber}},
  \bibinfo {author} {\bibfnamefont {J.}~\bibnamefont {Danckwerts}}, \ and\
  \bibinfo {author} {\bibfnamefont {A.}~\bibnamefont {Knorr}},\ }\href
  {\doibase 10.1103/PhysRevLett.91.127401} {\bibfield  {journal} {\bibinfo
  {journal} {Phys. Rev. Lett.}\ }\textbf {\bibinfo {volume} {91}},\ \bibinfo
  {pages} {127401} (\bibinfo {year} {2003})}\BibitemShut {NoStop}%
\bibitem [{\citenamefont {Carmele}\ \emph {et~al.}(2013)\citenamefont
  {Carmele}, \citenamefont {Knorr},\ and\ \citenamefont
  {Milde}}]{CarmeleMilde2013}%
  \BibitemOpen
  \bibfield  {author} {\bibinfo {author} {\bibfnamefont {A.}~\bibnamefont
  {Carmele}}, \bibinfo {author} {\bibfnamefont {A.}~\bibnamefont {Knorr}}, \
  and\ \bibinfo {author} {\bibfnamefont {F.}~\bibnamefont {Milde}},\ }\href
  {\doibase 10.1088/1367-2630/15/10/105024} {\bibfield  {journal} {\bibinfo
  {journal} {New Journal of Physics}\ }\textbf {\bibinfo {volume} {15}},\
  \bibinfo {pages} {105024} (\bibinfo {year} {2013})}\BibitemShut {NoStop}%
\bibitem [{\citenamefont {Chow}\ and\ \citenamefont {Jahnke}(2013)}]{Chow2013}%
  \BibitemOpen
  \bibfield  {author} {\bibinfo {author} {\bibfnamefont {W.~W.}\ \bibnamefont
  {Chow}}\ and\ \bibinfo {author} {\bibfnamefont {F.}~\bibnamefont {Jahnke}},\
  }\href {\doibase https://doi.org/10.1016/j.pquantelec.2013.04.001} {\bibfield
   {journal} {\bibinfo  {journal} {Progress in Quantum Electronics}\ }\textbf
  {\bibinfo {volume} {37}},\ \bibinfo {pages} {109 } (\bibinfo {year}
  {2013})}\BibitemShut {NoStop}%
\bibitem [{\citenamefont {Reiter}\ \emph {et~al.}(2014)\citenamefont {Reiter},
  \citenamefont {Kuhn}, \citenamefont {Glässl},\ and\ \citenamefont
  {Axt}}]{Reiter2014}%
  \BibitemOpen
  \bibfield  {author} {\bibinfo {author} {\bibfnamefont {D.~E.}\ \bibnamefont
  {Reiter}}, \bibinfo {author} {\bibfnamefont {T.}~\bibnamefont {Kuhn}},
  \bibinfo {author} {\bibfnamefont {M.}~\bibnamefont {Glässl}}, \ and\
  \bibinfo {author} {\bibfnamefont {V.~M.}\ \bibnamefont {Axt}},\ } {\bibfield  {journal} {\bibinfo
  {journal} {Journal of Physics: Condensed Matter}\ }\textbf {\bibinfo {volume}
  {26}},\ \bibinfo {pages} {423203} (\bibinfo {year} {2014})}\BibitemShut
  {NoStop}%
\bibitem [{\citenamefont {Reiter}\ \emph {et~al.}(2019)\citenamefont {Reiter},
  \citenamefont {Kuhn},\ and\ \citenamefont {Axt}}]{Reiter2019}%
  \BibitemOpen
  \bibfield  {author} {\bibinfo {author} {\bibfnamefont {D.~E.}\ \bibnamefont
  {Reiter}}, \bibinfo {author} {\bibfnamefont {T.}~\bibnamefont {Kuhn}}, \ and\
  \bibinfo {author} {\bibfnamefont {V.~M.}\ \bibnamefont {Axt}},\ }\href
  {\doibase 10.1080/23746149.2019.1655478} {\bibfield  {journal} {\bibinfo
  {journal} {Advances in Physics: X}\ }\textbf {\bibinfo {volume} {4}},\
  \bibinfo {pages} {1655478} (\bibinfo {year} {2019})}\BibitemShut {NoStop}%
\bibitem [{\citenamefont {Schilp}\ \emph {et~al.}(1994)\citenamefont {Schilp},
  \citenamefont {Kuhn},\ and\ \citenamefont {Mahler}}]{Schilp1994}%
  \BibitemOpen
  \bibfield  {author} {\bibinfo {author} {\bibfnamefont {J.}~\bibnamefont
  {Schilp}}, \bibinfo {author} {\bibfnamefont {T.}~\bibnamefont {Kuhn}}, \ and\
  \bibinfo {author} {\bibfnamefont {G.}~\bibnamefont {Mahler}},\ }\href
  {\doibase 10.1103/PhysRevB.50.5435} {\bibfield  {journal} {\bibinfo
  {journal} {Phys. Rev. B}\ }\textbf {\bibinfo {volume} {50}},\ \bibinfo
  {pages} {5435} (\bibinfo {year} {1994})}\BibitemShut {NoStop}%
\bibitem [{\citenamefont {Krummheuer}\ \emph {et~al.}(2002)\citenamefont
  {Krummheuer}, \citenamefont {Axt},\ and\ \citenamefont
  {Kuhn}}]{Krummheuer2002}%
  \BibitemOpen
  \bibfield  {author} {\bibinfo {author} {\bibfnamefont {B.}~\bibnamefont
  {Krummheuer}}, \bibinfo {author} {\bibfnamefont {V.~M.}\ \bibnamefont {Axt}},
  \ and\ \bibinfo {author} {\bibfnamefont {T.}~\bibnamefont {Kuhn}},\ }\href
  {\doibase 10.1103/PhysRevB.65.195313} {\bibfield  {journal} {\bibinfo
  {journal} {Phys. Rev. B}\ }\textbf {\bibinfo {volume} {65}},\ \bibinfo
  {pages} {195313} (\bibinfo {year} {2002})}\BibitemShut {NoStop}%
\bibitem [{\citenamefont {Rossi}\ and\ \citenamefont {Kuhn}(2002)}]{Rossi2002}%
  \BibitemOpen
  \bibfield  {author} {\bibinfo {author} {\bibfnamefont {F.}~\bibnamefont
  {Rossi}}\ and\ \bibinfo {author} {\bibfnamefont {T.}~\bibnamefont {Kuhn}},\
  }\href {\doibase 10.1103/RevModPhys.74.895} {\bibfield  {journal} {\bibinfo
  {journal} {Rev. Mod. Phys.}\ }\textbf {\bibinfo {volume} {74}},\ \bibinfo
  {pages} {895} (\bibinfo {year} {2002})}\BibitemShut {NoStop}%
\bibitem [{\citenamefont {Förstner}\ \emph {et~al.}(2003)\citenamefont
  {Förstner}, \citenamefont {Weber}, \citenamefont {Danckwerts},\ and\
  \citenamefont {Knorr}}]{Foerstner2003pssb}%
  \BibitemOpen
  \bibfield  {author} {\bibinfo {author} {\bibfnamefont {J.}~\bibnamefont
  {Förstner}}, \bibinfo {author} {\bibfnamefont {C.}~\bibnamefont {Weber}},
  \bibinfo {author} {\bibfnamefont {J.}~\bibnamefont {Danckwerts}}, \ and\
  \bibinfo {author} {\bibfnamefont {A.}~\bibnamefont {Knorr}},\ }\href
  {\doibase 10.1002/pssb.200303155} {\bibfield  {journal} {\bibinfo  {journal}
  {physica status solidi (b)}\ }\textbf {\bibinfo {volume} {238}},\ \bibinfo
  {pages} {419} (\bibinfo {year} {2003})}\BibitemShut {NoStop}%
\bibitem [{\citenamefont {Su}\ \emph {et~al.}(2013{\natexlab{b}})\citenamefont
  {Su}, \citenamefont {Bimberg}, \citenamefont {Knorr},\ and\ \citenamefont
  {Carmele}}]{su2013collective}%
  \BibitemOpen
  \bibfield  {author} {\bibinfo {author} {\bibfnamefont {Y.}~\bibnamefont
  {Su}}, \bibinfo {author} {\bibfnamefont {D.}~\bibnamefont {Bimberg}},
  \bibinfo {author} {\bibfnamefont {A.}~\bibnamefont {Knorr}}, \ and\ \bibinfo
  {author} {\bibfnamefont {A.}~\bibnamefont {Carmele}},\ }\href@noop {}
  {\bibfield  {journal} {\bibinfo  {journal} {Physical review letters}\
  }\textbf {\bibinfo {volume} {110}},\ \bibinfo {pages} {113604} (\bibinfo
  {year} {2013}{\natexlab{b}})}\BibitemShut {NoStop}%
\bibitem [{\citenamefont {Mukamel}(1999)}]{Mukamel1999}%
  \BibitemOpen
  \bibfield  {author} {\bibinfo {author} {\bibfnamefont {S.}~\bibnamefont
  {Mukamel}},\ }\href@noop {} {\emph {\bibinfo {title} {Principles of nonlinear
  optical spectroscopy}}}\ (\bibinfo  {publisher} {Oxford University Press},\
  \bibinfo {year} {1999})\BibitemShut {NoStop}%
\bibitem [{\citenamefont {Malic}\ and\ \citenamefont
  {Knorr}(2013)}]{Malic2013}%
  \BibitemOpen
  \bibfield  {author} {\bibinfo {author} {\bibfnamefont {E.}~\bibnamefont
  {Malic}}\ and\ \bibinfo {author} {\bibfnamefont {A.}~\bibnamefont {Knorr}},\
  }\href@noop {} {\emph {\bibinfo {title} {Graphene and carbon nanotubes:
  Ultrafast optics and relaxation dynamics}}}\ (\bibinfo  {publisher}
  {Wiley-VCH},\ \bibinfo {year} {2013})\BibitemShut {NoStop}%
\bibitem [{\citenamefont {Förstner}\ \emph {et~al.}(2002)\citenamefont
  {Förstner}, \citenamefont {Ahn}, \citenamefont {Danckwerts}, \citenamefont
  {Schaarschmidt}, \citenamefont {Waldmüller}, \citenamefont {Weber},\ and\
  \citenamefont {Knorr}}]{Foerstner2002}%
  \BibitemOpen
  \bibfield  {author} {\bibinfo {author} {\bibfnamefont {J.}~\bibnamefont
  {Förstner}}, \bibinfo {author} {\bibfnamefont {K.}~\bibnamefont {Ahn}},
  \bibinfo {author} {\bibfnamefont {J.}~\bibnamefont {Danckwerts}}, \bibinfo
  {author} {\bibfnamefont {M.}~\bibnamefont {Schaarschmidt}}, \bibinfo {author}
  {\bibfnamefont {I.}~\bibnamefont {Waldmüller}}, \bibinfo {author}
  {\bibfnamefont {C.}~\bibnamefont {Weber}}, \ and\ \bibinfo {author}
  {\bibfnamefont {A.}~\bibnamefont {Knorr}},\ }\href@noop {} {\bibfield
  {journal} {\bibinfo  {journal} {physica status solidi (b)}\ }\textbf
  {\bibinfo {volume} {234}},\ \bibinfo {pages} {155} (\bibinfo {year}
  {2002})}\BibitemShut {NoStop}%
\bibitem [{\citenamefont {Dexter}(1953)}]{Dexter1953}%
  \BibitemOpen
  \bibfield  {author} {\bibinfo {author} {\bibfnamefont {D.~L.}\ \bibnamefont
  {Dexter}},\ } {\bibfield  {journal}
  {\bibinfo  {journal} {The Journal of Chemical Physics}\ }\textbf {\bibinfo
  {volume} {21}},\ \bibinfo {pages} {836} (\bibinfo {year} {1953})}\BibitemShut
  {NoStop}%
\bibitem [{\citenamefont {Brusaferri}\ \emph {et~al.}(1996)\citenamefont
  {Brusaferri}, \citenamefont {Sanguinetti}, \citenamefont {Grilli},
  \citenamefont {Guzzi}, \citenamefont {Bignazzi}, \citenamefont {Bogani},
  \citenamefont {Carraresi}, \citenamefont {Colocci}, \citenamefont {Bosacchi},
  \citenamefont {Frigeri},\ and\ \citenamefont {Franchi}}]{Brusaferri1996}%
  \BibitemOpen
  \bibfield  {author} {\bibinfo {author} {\bibfnamefont {L.}~\bibnamefont
  {Brusaferri}}, \bibinfo {author} {\bibfnamefont {S.}~\bibnamefont
  {Sanguinetti}}, \bibinfo {author} {\bibfnamefont {E.}~\bibnamefont {Grilli}},
  \bibinfo {author} {\bibfnamefont {M.}~\bibnamefont {Guzzi}}, \bibinfo
  {author} {\bibfnamefont {A.}~\bibnamefont {Bignazzi}}, \bibinfo {author}
  {\bibfnamefont {F.}~\bibnamefont {Bogani}}, \bibinfo {author} {\bibfnamefont
  {L.}~\bibnamefont {Carraresi}}, \bibinfo {author} {\bibfnamefont
  {M.}~\bibnamefont {Colocci}}, \bibinfo {author} {\bibfnamefont
  {A.}~\bibnamefont {Bosacchi}}, \bibinfo {author} {\bibfnamefont
  {P.}~\bibnamefont {Frigeri}}, \ and\ \bibinfo {author} {\bibfnamefont
  {S.}~\bibnamefont {Franchi}},\ }\href {\doibase 10.1063/1.117304} {\bibfield
  {journal} {\bibinfo  {journal} {Applied Physics Letters}\ }\textbf {\bibinfo
  {volume} {69}},\ \bibinfo {pages} {3354} (\bibinfo {year}
  {1996})}\BibitemShut {NoStop}%
\bibitem [{\citenamefont {Sanguinetti}\ \emph {et~al.}(2000)\citenamefont
  {Sanguinetti}, \citenamefont {Padovani}, \citenamefont {Gurioli},
  \citenamefont {Grilli}, \citenamefont {Guzzi}, \citenamefont {Vinattieri},
  \citenamefont {Colocci}, \citenamefont {Frigeri},\ and\ \citenamefont
  {Franchi}}]{Sanguinetti2000}%
  \BibitemOpen
  \bibfield  {author} {\bibinfo {author} {\bibfnamefont {S.}~\bibnamefont
  {Sanguinetti}}, \bibinfo {author} {\bibfnamefont {M.}~\bibnamefont
  {Padovani}}, \bibinfo {author} {\bibfnamefont {M.}~\bibnamefont {Gurioli}},
  \bibinfo {author} {\bibfnamefont {E.}~\bibnamefont {Grilli}}, \bibinfo
  {author} {\bibfnamefont {M.}~\bibnamefont {Guzzi}}, \bibinfo {author}
  {\bibfnamefont {A.}~\bibnamefont {Vinattieri}}, \bibinfo {author}
  {\bibfnamefont {M.}~\bibnamefont {Colocci}}, \bibinfo {author} {\bibfnamefont
  {P.}~\bibnamefont {Frigeri}}, \ and\ \bibinfo {author} {\bibfnamefont
  {S.}~\bibnamefont {Franchi}},\ }\href {\doibase 10.1063/1.1290385} {\bibfield
   {journal} {\bibinfo  {journal} {Applied Physics Letters}\ }\textbf {\bibinfo
  {volume} {77}},\ \bibinfo {pages} {1307} (\bibinfo {year}
  {2000})}\BibitemShut {NoStop}%
\bibitem [{\citenamefont {Tarasov}\ \emph {et~al.}(2000)\citenamefont
  {Tarasov}, \citenamefont {Mazur}, \citenamefont {Zhuchenko}, \citenamefont
  {Maa{\ss}dorf}, \citenamefont {Nickel}, \citenamefont {Tomm}, \citenamefont
  {Kissel}, \citenamefont {Walther},\ and\ \citenamefont
  {Masselink}}]{Tarasov2000}%
  \BibitemOpen
  \bibfield  {author} {\bibinfo {author} {\bibfnamefont {G.~G.}\ \bibnamefont
  {Tarasov}}, \bibinfo {author} {\bibfnamefont {Y.~I.}\ \bibnamefont {Mazur}},
  \bibinfo {author} {\bibfnamefont {Z.~Y.}\ \bibnamefont {Zhuchenko}}, \bibinfo
  {author} {\bibfnamefont {A.}~\bibnamefont {Maa{\ss}dorf}}, \bibinfo {author}
  {\bibfnamefont {D.}~\bibnamefont {Nickel}}, \bibinfo {author} {\bibfnamefont
  {J.~W.}\ \bibnamefont {Tomm}}, \bibinfo {author} {\bibfnamefont
  {H.}~\bibnamefont {Kissel}}, \bibinfo {author} {\bibfnamefont
  {C.}~\bibnamefont {Walther}}, \ and\ \bibinfo {author} {\bibfnamefont
  {W.~T.}\ \bibnamefont {Masselink}},\ }\href {\doibase 10.1063/1.1323516}
  {\bibfield  {journal} {\bibinfo  {journal} {Journal of Applied Physics}\
  }\textbf {\bibinfo {volume} {88}},\ \bibinfo {pages} {7162} (\bibinfo {year}
  {2000})}\BibitemShut {NoStop}%
\bibitem [{\citenamefont {Mazur}\ \emph {et~al.}(2002)\citenamefont {Mazur},
  \citenamefont {Wang}, \citenamefont {Wang}, \citenamefont {Salamo},
  \citenamefont {Xiao},\ and\ \citenamefont {Kissel}}]{Mazur2002}%
  \BibitemOpen
  \bibfield  {author} {\bibinfo {author} {\bibfnamefont {Y.~I.}\ \bibnamefont
  {Mazur}}, \bibinfo {author} {\bibfnamefont {X.}~\bibnamefont {Wang}},
  \bibinfo {author} {\bibfnamefont {Z.~M.}\ \bibnamefont {Wang}}, \bibinfo
  {author} {\bibfnamefont {G.~J.}\ \bibnamefont {Salamo}}, \bibinfo {author}
  {\bibfnamefont {M.}~\bibnamefont {Xiao}}, \ and\ \bibinfo {author}
  {\bibfnamefont {H.}~\bibnamefont {Kissel}},\ }\href {\doibase
  10.1063/1.1510157} {\bibfield  {journal} {\bibinfo  {journal} {Applied
  Physics Letters}\ }\textbf {\bibinfo {volume} {81}},\ \bibinfo {pages} {2469}
  (\bibinfo {year} {2002})}\BibitemShut {NoStop}%
\bibitem [{\citenamefont {Mazur}\ \emph {et~al.}(2005)\citenamefont {Mazur},
  \citenamefont {Wang}, \citenamefont {Tarasov}, \citenamefont {Xiao},
  \citenamefont {Salamo}, \citenamefont {Tomm}, \citenamefont {Talalaev},\ and\
  \citenamefont {Kissel}}]{Mazur2005}%
  \BibitemOpen
  \bibfield  {author} {\bibinfo {author} {\bibfnamefont {Y.~I.}\ \bibnamefont
  {Mazur}}, \bibinfo {author} {\bibfnamefont {Z.~M.}\ \bibnamefont {Wang}},
  \bibinfo {author} {\bibfnamefont {G.~G.}\ \bibnamefont {Tarasov}}, \bibinfo
  {author} {\bibfnamefont {M.}~\bibnamefont {Xiao}}, \bibinfo {author}
  {\bibfnamefont {G.~J.}\ \bibnamefont {Salamo}}, \bibinfo {author}
  {\bibfnamefont {J.~W.}\ \bibnamefont {Tomm}}, \bibinfo {author}
  {\bibfnamefont {V.}~\bibnamefont {Talalaev}}, \ and\ \bibinfo {author}
  {\bibfnamefont {H.}~\bibnamefont {Kissel}},\ }\href {\doibase
  10.1063/1.1861980} {\bibfield  {journal} {\bibinfo  {journal} {Applied
  Physics Letters}\ }\textbf {\bibinfo {volume} {86}},\ \bibinfo {pages}
  {063102} (\bibinfo {year} {2005})}\BibitemShut {NoStop}%
\bibitem [{\citenamefont {Mazur}\ \emph {et~al.}(2009)\citenamefont {Mazur},
  \citenamefont {Dorogan}, \citenamefont {Marega}, \citenamefont {Tarasov},
  \citenamefont {Cesar}, \citenamefont {Lopez-Richard}, \citenamefont
  {Marques},\ and\ \citenamefont {Salamo}}]{Mazur2009}%
  \BibitemOpen
  \bibfield  {author} {\bibinfo {author} {\bibfnamefont {Y.~I.}\ \bibnamefont
  {Mazur}}, \bibinfo {author} {\bibfnamefont {V.~G.}\ \bibnamefont {Dorogan}},
  \bibinfo {author} {\bibfnamefont {E.}~\bibnamefont {Marega}}, \bibinfo
  {author} {\bibfnamefont {G.~G.}\ \bibnamefont {Tarasov}}, \bibinfo {author}
  {\bibfnamefont {D.~F.}\ \bibnamefont {Cesar}}, \bibinfo {author}
  {\bibfnamefont {V.}~\bibnamefont {Lopez-Richard}}, \bibinfo {author}
  {\bibfnamefont {G.~E.}\ \bibnamefont {Marques}}, \ and\ \bibinfo {author}
  {\bibfnamefont {G.~J.}\ \bibnamefont {Salamo}},\ }\href {\doibase
  10.1063/1.3103312} {\bibfield  {journal} {\bibinfo  {journal} {Applied
  Physics Letters}\ }\textbf {\bibinfo {volume} {94}},\ \bibinfo {pages}
  {123112} (\bibinfo {year} {2009})}\BibitemShut {NoStop}%
\bibitem [{\citenamefont {Bhattacharyya}\ \emph {et~al.}(2012)\citenamefont
  {Bhattacharyya}, \citenamefont {Zybell}, \citenamefont {Winnerl},
  \citenamefont {Helm}, \citenamefont {Hopkinson}, \citenamefont {Wilson},\
  and\ \citenamefont {Schneider}}]{Bhattacharyya2012}%
  \BibitemOpen
  \bibfield  {author} {\bibinfo {author} {\bibfnamefont {J.}~\bibnamefont
  {Bhattacharyya}}, \bibinfo {author} {\bibfnamefont {S.}~\bibnamefont
  {Zybell}}, \bibinfo {author} {\bibfnamefont {S.}~\bibnamefont {Winnerl}},
  \bibinfo {author} {\bibfnamefont {M.}~\bibnamefont {Helm}}, \bibinfo {author}
  {\bibfnamefont {M.}~\bibnamefont {Hopkinson}}, \bibinfo {author}
  {\bibfnamefont {L.~R.}\ \bibnamefont {Wilson}}, \ and\ \bibinfo {author}
  {\bibfnamefont {H.}~\bibnamefont {Schneider}},\ }\href {\doibase
  10.1063/1.3701578} {\bibfield  {journal} {\bibinfo  {journal} {Applied
  Physics Letters}\ }\textbf {\bibinfo {volume} {100}},\ \bibinfo {pages}
  {152101} (\bibinfo {year} {2012})}\BibitemShut {NoStop}%
\bibitem [{\citenamefont {Specht}\ \emph {et~al.}(2015)\citenamefont {Specht},
  \citenamefont {Knorr},\ and\ \citenamefont {Richter}}]{Specht2015}%
  \BibitemOpen
  \bibfield  {author} {\bibinfo {author} {\bibfnamefont {J.~F.}\ \bibnamefont
  {Specht}}, \bibinfo {author} {\bibfnamefont {A.}~\bibnamefont {Knorr}}, \
  and\ \bibinfo {author} {\bibfnamefont {M.}~\bibnamefont {Richter}},\ }\href
  {\doibase 10.1103/physrevb.91.155313} {\bibfield  {journal} {\bibinfo
  {journal} {Physical Review B}\ }\textbf {\bibinfo {volume} {91}},\ \bibinfo
  {pages} {155313} (\bibinfo {year} {2015})}\BibitemShut {NoStop}%
\bibitem [{\citenamefont {Förster}(1948)}]{Foerster1948}%
  \BibitemOpen
  \bibfield  {author} {\bibinfo {author} {\bibfnamefont {T.}~\bibnamefont
  {Förster}},\ }\href {\doibase 10.1002/andp.19484370105} {\bibfield
  {journal} {\bibinfo  {journal} {Ann. Phys.}\ }\textbf {\bibinfo {volume}
  {437}},\ \bibinfo {pages} {55} (\bibinfo {year} {1948})}\BibitemShut
  {NoStop}%
\bibitem [{\citenamefont {Clapp}\ \emph {et~al.}(2006)\citenamefont {Clapp},
  \citenamefont {Medintz},\ and\ \citenamefont {Mattoussi}}]{Clapp2006}%
  \BibitemOpen
  \bibfield  {author} {\bibinfo {author} {\bibfnamefont {A.~R.}\ \bibnamefont
  {Clapp}}, \bibinfo {author} {\bibfnamefont {I.~L.}\ \bibnamefont {Medintz}},
  \ and\ \bibinfo {author} {\bibfnamefont {H.}~\bibnamefont {Mattoussi}},\
  }\href {\doibase 10.1002/cphc.200500217} {\bibfield  {journal} {\bibinfo
  {journal} {ChemPhysChem}\ }\textbf {\bibinfo {volume} {7}},\ \bibinfo {pages}
  {47} (\bibinfo {year} {2006})}\BibitemShut {NoStop}%
\bibitem [{\citenamefont {Rozbicki}\ and\ \citenamefont
  {Machnikowski}(2008)}]{Rozbicki2008}%
  \BibitemOpen
  \bibfield  {author} {\bibinfo {author} {\bibfnamefont {E.}~\bibnamefont
  {Rozbicki}}\ and\ \bibinfo {author} {\bibfnamefont {P.}~\bibnamefont
  {Machnikowski}},\ }\href {\doibase 10.1103/PhysRevLett.100.027401} {\bibfield
   {journal} {\bibinfo  {journal} {Phys. Rev. Lett.}\ }\textbf {\bibinfo
  {volume} {100}},\ \bibinfo {pages} {027401} (\bibinfo {year}
  {2008})}\BibitemShut {NoStop}%
\bibitem [{\citenamefont {Machnikowski}\ and\ \citenamefont
  {Rozbicki}(2009)}]{Machnikowski2009}%
  \BibitemOpen
  \bibfield  {author} {\bibinfo {author} {\bibfnamefont {P.}~\bibnamefont
  {Machnikowski}}\ and\ \bibinfo {author} {\bibfnamefont {E.}~\bibnamefont
  {Rozbicki}},\ }\href {\doibase 10.1002/pssb.200880312} {\bibfield  {journal}
  {\bibinfo  {journal} {physica status solidi (b)}\ }\textbf {\bibinfo {volume}
  {246}},\ \bibinfo {pages} {320} (\bibinfo {year} {2009})}\BibitemShut
  {NoStop}%
\bibitem [{\citenamefont {Yuan}\ \emph {et~al.}(2015)\citenamefont {Yuan},
  \citenamefont {Xu},\ and\ \citenamefont {Fan}}]{Yuan2015}%
  \BibitemOpen
  \bibfield  {author} {\bibinfo {author} {\bibfnamefont {L.}~\bibnamefont
  {Yuan}}, \bibinfo {author} {\bibfnamefont {S.}~\bibnamefont {Xu}}, \ and\
  \bibinfo {author} {\bibfnamefont {S.}~\bibnamefont {Fan}},\ }\href {\doibase
  10.1364/OL.40.005140} {\bibfield  {journal} {\bibinfo  {journal} {Opt.
  Lett.}\ }\textbf {\bibinfo {volume} {40}},\ \bibinfo {pages} {5140} (\bibinfo
  {year} {2015})}\BibitemShut {NoStop}%
\bibitem [{\citenamefont {Lin}\ and\ \citenamefont {Fan}(2014)}]{Lin2014}%
  \BibitemOpen
  \bibfield  {author} {\bibinfo {author} {\bibfnamefont {Q.}~\bibnamefont
  {Lin}}\ and\ \bibinfo {author} {\bibfnamefont {S.}~\bibnamefont {Fan}},\
  }\href {\doibase 10.1103/PhysRevX.4.031031} {\bibfield  {journal} {\bibinfo
  {journal} {Phys. Rev. X}\ }\textbf {\bibinfo {volume} {4}},\ \bibinfo {pages}
  {031031} (\bibinfo {year} {2014})}\BibitemShut {NoStop}%
\bibitem [{\citenamefont {Peano}\ \emph {et~al.}(2015)\citenamefont {Peano},
  \citenamefont {Brendel}, \citenamefont {Schmidt},\ and\ \citenamefont
  {Marquardt}}]{Peano2015}%
  \BibitemOpen
  \bibfield  {author} {\bibinfo {author} {\bibfnamefont {V.}~\bibnamefont
  {Peano}}, \bibinfo {author} {\bibfnamefont {C.}~\bibnamefont {Brendel}},
  \bibinfo {author} {\bibfnamefont {M.}~\bibnamefont {Schmidt}}, \ and\
  \bibinfo {author} {\bibfnamefont {F.}~\bibnamefont {Marquardt}},\ }\href
  {\doibase 10.1103/PhysRevX.5.031011} {\bibfield  {journal} {\bibinfo
  {journal} {Phys. Rev. X}\ }\textbf {\bibinfo {volume} {5}},\ \bibinfo {pages}
  {031011} (\bibinfo {year} {2015})}\BibitemShut {NoStop}%
\bibitem [{\citenamefont {Peano}\ \emph {et~al.}(2016)\citenamefont {Peano},
  \citenamefont {Houde}, \citenamefont {Marquardt},\ and\ \citenamefont
  {Clerk}}]{Peano2016}%
  \BibitemOpen
  \bibfield  {author} {\bibinfo {author} {\bibfnamefont {V.}~\bibnamefont
  {Peano}}, \bibinfo {author} {\bibfnamefont {M.}~\bibnamefont {Houde}},
  \bibinfo {author} {\bibfnamefont {F.}~\bibnamefont {Marquardt}}, \ and\
  \bibinfo {author} {\bibfnamefont {A.~A.}\ \bibnamefont {Clerk}},\ }\href
  {\doibase 10.1103/PhysRevX.6.041026} {\bibfield  {journal} {\bibinfo
  {journal} {Phys. Rev. X}\ }\textbf {\bibinfo {volume} {6}},\ \bibinfo {pages}
  {041026} (\bibinfo {year} {2016})}\BibitemShut {NoStop}%
\bibitem [{\citenamefont {Bandres}\ \emph {et~al.}(2016)\citenamefont
  {Bandres}, \citenamefont {Rechtsman},\ and\ \citenamefont
  {Segev}}]{Bandres2016}%
  \BibitemOpen
  \bibfield  {author} {\bibinfo {author} {\bibfnamefont {M.~A.}\ \bibnamefont
  {Bandres}}, \bibinfo {author} {\bibfnamefont {M.~C.}\ \bibnamefont
  {Rechtsman}}, \ and\ \bibinfo {author} {\bibfnamefont {M.}~\bibnamefont
  {Segev}},\ }\href {\doibase 10.1103/PhysRevX.6.011016} {\bibfield  {journal}
  {\bibinfo  {journal} {Phys. Rev. X}\ }\textbf {\bibinfo {volume} {6}},\
  \bibinfo {pages} {011016} (\bibinfo {year} {2016})}\BibitemShut {NoStop}%
\bibitem [{\citenamefont {Poshakinskiy}\ and\ \citenamefont
  {Poddubny}(2019)}]{Posha2019}%
  \BibitemOpen
  \bibfield  {author} {\bibinfo {author} {\bibfnamefont {A.~V.}\ \bibnamefont
  {Poshakinskiy}}\ and\ \bibinfo {author} {\bibfnamefont {A.~N.}\ \bibnamefont
  {Poddubny}},\ }\href {\doibase 10.1103/PhysRevX.9.011008} {\bibfield
  {journal} {\bibinfo  {journal} {Phys. Rev. X}\ }\textbf {\bibinfo {volume}
  {9}},\ \bibinfo {pages} {011008} (\bibinfo {year} {2019})}\BibitemShut
  {NoStop}%
\bibitem [{\citenamefont {Mahmoodian}\ \emph {et~al.}(2020)\citenamefont
  {Mahmoodian}, \citenamefont {Calaj\'o}, \citenamefont {Chang}, \citenamefont
  {Hammerer},\ and\ \citenamefont {S\o{}rensen}}]{Mahmo2020}%
  \BibitemOpen
  \bibfield  {author} {\bibinfo {author} {\bibfnamefont {S.}~\bibnamefont
  {Mahmoodian}}, \bibinfo {author} {\bibfnamefont {G.}~\bibnamefont
  {Calaj\'o}}, \bibinfo {author} {\bibfnamefont {D.~E.}\ \bibnamefont {Chang}},
  \bibinfo {author} {\bibfnamefont {K.}~\bibnamefont {Hammerer}}, \ and\
  \bibinfo {author} {\bibfnamefont {A.~S.}\ \bibnamefont {S\o{}rensen}},\
  }\href {\doibase 10.1103/PhysRevX.10.031011} {\bibfield  {journal} {\bibinfo
  {journal} {Phys. Rev. X}\ }\textbf {\bibinfo {volume} {10}},\ \bibinfo
  {pages} {031011} (\bibinfo {year} {2020})}\BibitemShut {NoStop}%
\bibitem [{\citenamefont {Kawazoe}\ \emph {et~al.}(2002)\citenamefont
  {Kawazoe}, \citenamefont {Kobayashi}, \citenamefont {Lim}, \citenamefont
  {Narita},\ and\ \citenamefont {Ohtsu}}]{Kawazoe2002}%
  \BibitemOpen
  \bibfield  {author} {\bibinfo {author} {\bibfnamefont {T.}~\bibnamefont
  {Kawazoe}}, \bibinfo {author} {\bibfnamefont {K.}~\bibnamefont {Kobayashi}},
  \bibinfo {author} {\bibfnamefont {J.}~\bibnamefont {Lim}}, \bibinfo {author}
  {\bibfnamefont {Y.}~\bibnamefont {Narita}}, \ and\ \bibinfo {author}
  {\bibfnamefont {M.}~\bibnamefont {Ohtsu}},\ }\href {\doibase
  10.1103/physrevlett.88.067404} {\bibfield  {journal} {\bibinfo  {journal}
  {Phys. Rev. Lett.}\ }\textbf {\bibinfo {volume} {88}},\ \bibinfo {pages}
  {067404} (\bibinfo {year} {2002})}\BibitemShut {NoStop}%
\bibitem [{\citenamefont {Kobayashi}\ \emph {et~al.}(2003)\citenamefont
  {Kobayashi}, \citenamefont {Sangu}, \citenamefont {Shojiguchi}, \citenamefont
  {Kawazoe}, \citenamefont {Kitahara},\ and\ \citenamefont
  {Ohtsu}}]{Kobayashi2003}%
  \BibitemOpen
  \bibfield  {author} {\bibinfo {author} {\bibfnamefont {K.}~\bibnamefont
  {Kobayashi}}, \bibinfo {author} {\bibfnamefont {S.}~\bibnamefont {Sangu}},
  \bibinfo {author} {\bibfnamefont {A.}~\bibnamefont {Shojiguchi}}, \bibinfo
  {author} {\bibfnamefont {T.}~\bibnamefont {Kawazoe}}, \bibinfo {author}
  {\bibfnamefont {K.}~\bibnamefont {Kitahara}}, \ and\ \bibinfo {author}
  {\bibfnamefont {M.}~\bibnamefont {Ohtsu}},\ }\href {\doibase
  10.1046/j.1365-2818.2003.01131.x} {\bibfield  {journal} {\bibinfo  {journal}
  {Journal of Microscopy}\ }\textbf {\bibinfo {volume} {210}},\ \bibinfo
  {pages} {247} (\bibinfo {year} {2003})}\BibitemShut {NoStop}%
\bibitem [{\citenamefont {Sangu}\ \emph {et~al.}(2004)\citenamefont {Sangu},
  \citenamefont {Kobayashi}, \citenamefont {Shojiguchi},\ and\ \citenamefont
  {Ohtsu}}]{Sangu2004}%
  \BibitemOpen
  \bibfield  {author} {\bibinfo {author} {\bibfnamefont {S.}~\bibnamefont
  {Sangu}}, \bibinfo {author} {\bibfnamefont {K.}~\bibnamefont {Kobayashi}},
  \bibinfo {author} {\bibfnamefont {A.}~\bibnamefont {Shojiguchi}}, \ and\
  \bibinfo {author} {\bibfnamefont {M.}~\bibnamefont {Ohtsu}},\ }\href
  {\doibase 10.1103/physrevb.69.115334} {\bibfield  {journal} {\bibinfo
  {journal} {Physical Review B}\ }\textbf {\bibinfo {volume} {69}},\ \bibinfo
  {pages} {115334} (\bibinfo {year} {2004})}\BibitemShut {NoStop}%
\bibitem [{\citenamefont {Sato}\ \emph {et~al.}(2007)\citenamefont {Sato},
  \citenamefont {Minami},\ and\ \citenamefont {Kobayashi}}]{Sato2007}%
  \BibitemOpen
  \bibfield  {author} {\bibinfo {author} {\bibfnamefont {A.}~\bibnamefont
  {Sato}}, \bibinfo {author} {\bibfnamefont {F.}~\bibnamefont {Minami}}, \ and\
  \bibinfo {author} {\bibfnamefont {K.}~\bibnamefont {Kobayashi}},\ } {\bibfield  {journal} {\bibinfo
  {journal} {Physica E: Low-dimensional Systems and Nanostructures}\ }\textbf
  {\bibinfo {volume} {40}},\ \bibinfo {pages} {313} (\bibinfo {year}
  {2007})}\BibitemShut {NoStop}%
\bibitem [{\citenamefont {Nishibayashi}\ \emph {et~al.}(2008)\citenamefont
  {Nishibayashi}, \citenamefont {Kawazoe}, \citenamefont {Ohtsu}, \citenamefont
  {Akahane},\ and\ \citenamefont {Yamamoto}}]{Nishibayashi2008}%
  \BibitemOpen
  \bibfield  {author} {\bibinfo {author} {\bibfnamefont {K.}~\bibnamefont
  {Nishibayashi}}, \bibinfo {author} {\bibfnamefont {T.}~\bibnamefont
  {Kawazoe}}, \bibinfo {author} {\bibfnamefont {M.}~\bibnamefont {Ohtsu}},
  \bibinfo {author} {\bibfnamefont {K.}~\bibnamefont {Akahane}}, \ and\
  \bibinfo {author} {\bibfnamefont {N.}~\bibnamefont {Yamamoto}},\ }\href
  {\doibase 10.1063/1.2945289} {\bibfield  {journal} {\bibinfo  {journal}
  {Applied Physics Letters}\ }\textbf {\bibinfo {volume} {93}},\ \bibinfo
  {pages} {042101} (\bibinfo {year} {2008})}\BibitemShut {NoStop}%
\end{thebibliography}

%

\end{document}